\documentclass[sn-aps]{sn-jnl}

\usepackage{graphicx}%
\usepackage{multirow}%
\usepackage{amsmath,amssymb,amsfonts}%
\usepackage{amsthm}%
\usepackage{mathrsfs}%
\usepackage[title]{appendix}%
\usepackage{xcolor}%
\usepackage{textcomp}%
\usepackage{manyfoot}%
\usepackage{booktabs}%
\usepackage{algorithm}%
\usepackage{algorithmicx}%
\usepackage{algpseudocode}%
\usepackage{listings}%

\usepackage{mathtools}
\usepackage{tabularx}
\usepackage{makecell}

\usepackage{subcaption}

\usepackage[T1]{fontenc}
\usepackage{pifont}
\usepackage{colortbl}

\usepackage{comment}

\let\orcidlogo\relax
\usepackage{orcidlink}

\newcommand*\rot{\rotatebox{90}}

\begin{document}

\title[Article Title]{BiCoRec: Bias-Mitigated Context-Aware Sequential Recommendation Model}

\author*[1]{\fnm{Mufhumudzi} \sur{Muthivhi} \orcidlink{0000-0003-0509-6235}}\email{15mmuthivhi@gmail.com}

\author[2]{\fnm{Terence L.} \sur{van Zyl} \orcidlink{0000-0003-4281-630X}}\email{tvanzyl@uj.ac.za}

\author[3]{\fnm{Hairong} \sur{Wang} \orcidlink{0000-0001-8770-5916}}\email{hairongwng@gmail.com}

\affil*[1, 3]{\orgname{University of Witwatersrand}, \orgaddress{\city{Johannesburg}, \postcode{2090}, \state{GT}, \country{South Africa}}}

\affil[2]{\orgname{University of Johannesburg}, \orgaddress{\city{Johannesburg}, \postcode{2092}, \state{GT}, \country{South Africa}}}


\abstract{
Sequential recommendation models aim to learn from users' evolving preferences. 
However, current state-of-the-art models suffer from an inherent popularity bias.
This study developed a novel framework, BiCoRec, that adaptively accommodates users' changing preferences for popular and niche items.
Our approach leverages a co-attention mechanism to obtain a popularity-weighted user sequence representation, facilitating more accurate predictions. We then present a new training scheme that learns from future preferences using a consistency loss function.
BiCoRec aimed to improve the recommendation performance of users who preferred niche items. For these users, BiCoRec achieves a 26.00\% average improvement in NDCG@10 over state-of-the-art baselines. When ranking the relevant item against the entire collection, BiCoRec achieves NDCG@10 scores of 0.0102, 0.0047, 0.0021, and 0.0005 for the Movies, Fashion, Games and Music datasets.}




\maketitle

\section{Introduction}\label{Introduction}

Recommendation models help alleviate the information overload problem~\cite{billsus1998learning}. They are tasked with suggesting a relevant item from a larger collection of items. Some of the earliest models used a Collaborative Filtering (CF) approach. CF predicts the item that a user would prefer by observing the interests of similar users~\cite{goldberg1992using}. Initially, Matrix Factorization was applied to project the user and item features into the same latent factor space. However, CF-based models disregard the sequential dependencies between user interactions.

Sequential recommendation models represent the interacted items in sequential order~\cite{zimdars2013using, shani2005mdp, he2017translation, wang2015learning, tang2018personalized, jannach2017recurrent}. Their objective is to predict the next item that the user will prefer, based on their historical interactions. Research has successfully incorporated past interactions using Markov-Chains (MC), Convolutional Neural Networks (CNN), Recurrent Neural Networks (RNN), and attention-based mechanisms.

Contemporary work has prioritized enhancing the expressive power of the learned item features~\cite{hou2022towards, geng2023vip5}. They consider any contextual information that describes the visual, textual, or auditory attributes of the item. These multi-modal signals enrich the model's understanding of the item.

Despite current progress, sequential recommendation models struggle with \textit{popularity bias}~\cite{yang2023debiased, muthivhi2023impacts}. The frequency of items in a recommendation dataset follows a power law distribution~\cite{abdollahpouri2019managing}. A small number of items occupy the short head of the power law probability distribution~\cite{park2008long}. The remaining majority of items occupy the long tail of the distribution. The short-head (\textit{popular}) and long-tailed (\textit{niche}) items produce an inherent bias within the dataset where a small number of items contribute towards a substantial number of interactions~\cite{abdollahpouri2017controlling}. This results in users being exposed to popular items more frequently than their popularity would warrant~\cite{abdollahpouri2020multi, abdollahpouri2017controlling}.

Several investigations have focused on addressing popularity bias~\cite{abdollahpouri2017controlling, adomavicius2011improving}. Nonetheless, only a few studies have incorporated bias mitigation techniques within the context of sequential recommendations~\cite{yang2023debiased, muthivhi2023impacts}. The study found that users with long sequences, where the items are arranged from past to present, shift their preferences towards more niche items over time. Intuitively, their tastes refine towards a distinct set of items that caters to their unique arrangement. To mitigate the popularity bias in multi-modal sequential recommendation models, we propose \underline{\textbf{Bi}}as-Mitigated \underline{\textbf{Co}}ntext-Aware Sequential \underline{\textbf{Rec}}ommendation Model (BiCoRec) that adaptively accommodates the evolving preferences of users. The empirical results show that BiCoRec, on average, is $7 \%$ more capable of uncovering the most relevant items compared to the baseline models. The proposed bias mitigation techniques in BiCoRec improve the performance of users who prefer less popular items by $26 \%$. Further, BiCoRec achieves an average of $3.14 \%$ increase in the NDCG score for long user sequence datasets over current SOTA models.

\subsection{Contributions}

In Section~\ref{section:ProblemSetting}, we characterize popularity bias in sequential recommendation by analyzing the sequence of items preceding the target item. We refer to this sequence of items as the \textit{history} of user preferences. Our empirical analysis reveals two key findings:
\begin{enumerate}
    \item current sequential recommendation models struggle to learn robust representations for items with sparse historical interactions; and
    \item users increasingly prefer niche items over time. Hence, this exacerbates the problem, as niche items have limited interaction histories.
\end{enumerate}
BiCoRec addresses the above problems by making three key contributions:
\begin{itemize}
    \item Popularity-Aware Embedding
        \begin{itemize}
            \item Generates a popularity embedding that explicitly encodes the bias inherent in the user's sequence if items.
        \end{itemize}
    \item Co-Attention-Based Weighting
        \begin{itemize}
            \item Uses this embedding to dynamically re-weight items in the user’s sequence via a co-attention mechanism.
        \end{itemize}
    \item Future Preference Anticipation
        \begin{itemize}
            \item Adopt a semi-supervised paradigm, cross-pseudo supervision, to predict a sequence of future items rather than just the next item. This method learns from both labelled observed items and unlabeled padded items.
        \end{itemize}
\end{itemize}
Finally, we provide a novel evaluation scheme that measures users' evolving preferences for the sequential recommendation domain.

The remainder of the paper is structured as follows: Section~\ref{section:Background} provides a brief background in Recommendation Systems. In Section~\ref{section:ProblemSetting}, we outline two distinct problems addressed in this paper. In Section~\ref{section:Model}, we present the proposed model framework. Section~\ref{section:Methodology} discusses the methodology, while Section~\ref{section:Results} presents our results and discussion. Finally, Section~\ref{section:Conclusion} concludes by summarising its contributions and presenting recommendations for future work.

\section{Background}\label{section:Background}

\subsection{General Recommendation} 

Some early recommendation models used Collaborative Filtering (CF)~\cite{goldberg1992using}. They learn user preferences by encoding the collaborative signal within the interaction data. The most successful CF-based models adopt a representation learning framework~\cite{Koren2009Matrix}. This process, known as Matrix Factorisation (MF), is tasked with identifying the latent spaces that encode the user-item interaction matrix. However, the linear inner product MF operation may not be sufficient to model the complex structure of users' interactions. In response to this inflexibility, He \textit{et al.}~\cite{he2017neural} apply a non-linear continuous function to capture intricate dependencies between user and item interactions. Their Neural Collaborative Filtering (NCF) model uses a multi-layer neural network to approximate any continuous function. MF follows a point-wise learning regression framework that minimizes the squared loss between the predicted value and the target value. Bayesian Personalized Ranking (BPR) adopts a pair-wise ranking-based method~\cite{rendle2012bpr}. More specifically, it aims to correctly rank item pairs rather than scoring individual items. It finds an accurate personalized ranking for all items by optimizing the maximum posterior probability. Collaborative filtering assumes that the users prefer items that other users have already interacted with. Any new or unobserved items would not get recommended.
\subsection{Sequential Recommendation} 

Zimdars \textit{et al.}~\cite{zimdars2013using} extend collaborative filtering by representing the interacted items in sequential order such that the model becomes a prediction problem. Given a user’s past actions, we can estimate their future actions by modelling transitions in the sequences of items. The model parameters now encode the evolution of user preferences, which may be independent of those of other users. 
Shani \textit{et al.}~\cite{shani2005mdp} approach the problem by using a Markov Chain (MC) to make a decision at each time step. An MC captures sequential effects over time using a transition matrix. He \textit{et al.}~\cite{he2017translation} propose a first-order Markov chain using a single vector that captures the correlation between preferences and sequential continuity. Their \underline{Trans}lation-based \underline{Rec}ommender (TransRec) model finds the latent translation vector that encodes a user's inherent intent or ``long-term preferences'' that influenced their decision making. Tang and Wang \textit{et al.}~\cite{tang2018personalized} further expand the sequential decision problem by applying a high-order MC. They argue that sequences reflect a union-level influence where several previous items, in that order, jointly influence the target item. Their proposed \underline{C}onvolution\underline{A}l \underline{S}equence \underline{E}mbedding \underline{R}ecommendation \underline{M}odel (Caser) adopts a deep learning-based Convolutional Neural Network (CNN) to learn sequential features. The horizontal and vertical convolutional filters produce an embedding matrix as an image of the previous items in the latent space by searching for sequential patterns as local features. Jannach and
Ludewig \textit{et al.}~\cite{jannach2017recurrent} achieve a similar effect to Caser without a convolutional neural network. They use a Recurrent Neural Network (RNN) and take the sequence of items as input and efficiently predict the next item. Their \underline{G}ated \underline{R}ecurrent \underline{U}nit \underline{for} Sequential \underline{Rec}ommendation (GRU4Rec) model is further enhanced with a pairwise learning-to-rank module. Similar to BPR, GRU4Rec learning objective is to rank the relevant item higher within the list. 

However, RNNs require a large amount of data to train. They are susceptible to noise because they assume that the entire sequence of items is necessary for predicting the next item. Conversely, first-order (TransRec) and high-order (Caser) Markov Chains benefit from inferring future preference from single or multiple past actions. On the other hand, the attention mechanism aims to learn from all past actions (RNNs) whilst inferring predictions from a smaller subset of actions (MCs). Attention mechanisms adaptively assign weights to previous items at each time step~\cite{kang2018self}. The model places importance on the subset
of items relevant to the next observed item. \underline{S}elf-\underline{A}ttentive \underline{S}equential \underline{Rec}ommander (SASRec) uses a ``self-attention'' module to encode the user's historical interactions uni-directionally from left to right. Each item can only encode the information from previous items. In contrast, Sun \textit{et al.} ~\cite{sun2019bert4rec} argue that a unidirectional transformer is insufficient. They propose a bi-directional learning architecture that incorporates context from both left-to-right and right-to-left directions. The \underline{B}idirectional \underline{E}ncoder \underline{R}epresentation \underline{T}ransformer (BERT) uses a Multi-Head Attention (MH) mechanism to incorporate context from not only past items but also future observations~\cite{sun2019bert4rec}. 

\subsection{Context-Aware Recommendation} 

The above temporal-based models aim only to encode the context of time in relation to user preferences. However, context is any information that characterizes a situation related to the interaction between humans, applications and the surrounding environment~\cite{bazire2005understanding}. Auxiliary information is any additional insights about an item or user beyond the interaction data. Factorization Machines (FMs) use item attributes from tabular data as auxiliary information~\cite{rendle2010factorization}. The item attribute data may include the name, colour or size of the item. Instead of the original user-item interaction matrix, FM produces a description matrix from the item attributes data. 

Fischer \textit{et al.}~\cite{fischer2020integrating} concatenate the item attributes to the item embedding and then trains on a bidirectional transformer. The tabular data is fed into the transformer by one-hot encoding the keyword descriptions. Zhang \textit{et al.}~\cite{zhang2019feature} extend the item attributes to text descriptions and employs two separate self-attention blocks for item attribute data and text descriptions. The text description provides further context about the item in an unstructured format. Singer \textit{et al.}~\cite{singer2022sequential} track the changes of item attributes over time through a transformer that handles 2D input sequences. Zhou \textit{et al.}~\cite{zhou2020s3} design a contrastive learning loss function to maximize the mutual information between items and attributes. \underline{Ke}yword \underline{BERT4Rec} (KeBERT4Rec) extends on BERT4Rec by including an auxiliary information embedding~\cite{fischer2020integrating}. The auxiliary information embedding is obtained from encoding categorical data into a multi-hot encoded vector.

\subsection{Bias-Mitigation} 

Some of the earliest methods proposed a De-biased Matrix Factorization strategy to alleviate popularity bias~\cite{Koren2009Matrix}. This method encapsulates the effects of bias, which do not involve user-item interaction, within a baseline predictor. Several subsequent investigations have focused on addressing popularity bias for traditional recommendation models~\cite{abdollahpouri2017controlling, adomavicius2011improving, zhu2021popularity, chen2020esam, zheng2021disentangling}. Nonetheless, only a few studies have incorporated bias mitigation techniques within the context of sequential recommendations~\cite{yang2023debiased, wu2024popularity}. Sequential data often consists of complicated dependencies, of which the debiasing methods used for traditional recommendation cannot be applied directly~\cite{wu2024popularity}. Yang \textit{et al.}~\cite{yang2023debiased} mitigate popularity bias by examining the data augmentation process for contrastive learning in sequential recommendation. They extract
self-supervision signals with conformity and interest disentanglement to learn augmented representations aware of popularity bias. Wu \textit{et al.}~\cite{wu2024popularity} argue that the next item prediction is influenced by popularity. They identify the causal effect of popularity through a graph and then consider the user's desire that affects the effect of popularity.

\subsection{Conclusion} 

The current state-of-the-art recommendation models enhance collaborative filtering by arranging items in sequential order. Deep-learning-based sequential recommendation models successfully encode each user's preferences into a latent representation. Context-aware recommendation improves the representation space by including item auxiliary information. Popularity bias-mitigation methods aim to reduce the effects of popularity bias to improve the diversity of recommended items. Despite progress in context-aware and bias-mitigation recommendation systems, few studies integrate both aspects in the sequential recommendation setting. In the next section, we formalize the problem of popularity bias in sequential recommendation models.

\section{Problem Setting}\label{section:ProblemSetting}

The problem addressed in this study is formulated as follows. Given a set of users $\mathcal{U} = \{u_1, u_2, \dots, u_{|\mathcal{U}|}\}$, a set of items by $\mathcal{V} = \{v_1, v_2, \dots, v_{|\mathcal{V}|}\}$, the auxiliary information data $\mathcal{A} = \{ a_{1}, a_{2}, \dots, a_{|\mathcal{V}|} \}$, the user sequence popularity scores $\mathbf{Q} \in \mathbb{R}^{\vert \mathcal{U} \vert \times n}$ and the sequential item data $\mathcal{S}_u = \{v_1^u, v_2^u, \dots, v_{n}^u\}$ where $v_t^u \in \mathcal{V}$ represents an interaction of user $u$ with item $v$ at relative time $t$ and $n$ is the maximum sequence length, predict the next item $v_{n + 1}^u$ that is most preferred by user $u$:
\begin{equation}
    p(v_{n + 1}^u \vert \mathcal{S}_u, \mathcal{A}, \mathbf{q}_{u})
\end{equation}
which is the probability over all possible items for user $u$ at time step $n + 1$ and $\mathbf{q}_{u} \in \mathbf{Q}$~\cite{muthivhi2022multi,muthivhi2023impacts,kang2018self,sun2019bert4rec}.

\subsection{Popularity Bias}

Sequential models aim to learn the latent features of the item $v_{t}^{u}$ based on the \textit{history} of items $v_{1:t-1}^{u} = \{ v_{1}^{u}, v_{2}^{u}, \dots, v_{t-1}^{u} \}$ that precedes it. Specifically, the history is a truncated sequential list of items just before the most recent item. Let $\mathcal{U}_{v}$ be the users interacting with the item $v$. Then user $u \in \mathcal{U}_{v}$ selected the item $v$ at time $t$ as well as the items $v_{1:t-1}$ that precedes the item $v_{t}^{u}$. Let $\mathcal{H}_{v} = \{ v_{1:t-1}^{u} \vert u \in \mathcal{U}_{v} \}$ be the set of \textit{histories} that precedes the item $v_t$ over all the users that have interacted with the item $v$. The set of histories is a truncated list of items for each user who has interacted with the item $v$. The task of sequential recommendation models is to learn the latent features of item $v$ based on its set of histories $\mathcal{H}_{v}$.

If $v_{\mathrm{pop}}$ is a popular item and $v_{\mathrm{niche}}$ is a niche item with fewer interactions, then:
\begin{equation}
    \vert \mathcal{H}_{v_{\mathrm{pop}}} \vert \gg \vert \mathcal{H}_{v_{\mathrm{niche}}} \vert
\end{equation}
that is the number of histories $\mathcal{H}_{v_{\mathrm{pop}}}$ that precede the popular item $v_{\mathrm{pop}}$ is always much greater than the number of histories that precede the niche item. Hence, the training data for a popular item is substantially greater than that of a niche or less popular item. 

\begin{figure}[htb!]
\centerline{\includegraphics[width=1.0\textwidth]{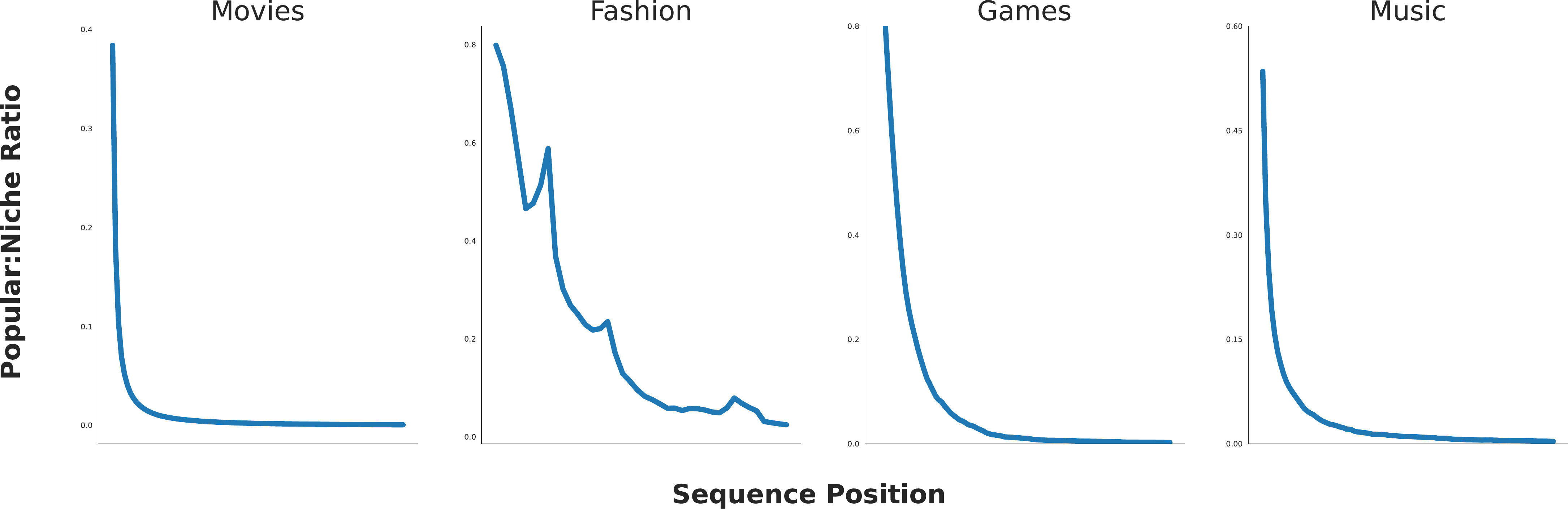}}
\caption{The ratio of popular to niche items within each position in the sequence. The ratio decreases as the sequence length increases, which shows users shift their preference from popular items to unpopular items over time}
\label{figure:Ratio}
\end{figure}

\subsection{Evolving Future Preferences} 

The analysis reveals that users with longer sequences tend to prefer less popular items over time. The frequency of items follows a power-law distribution. Following the approach outlined by Abdollahpouri \textit{et al.}~\cite{abdollahpouri2019managing}, this analysis selects the first 20\% of items from the power law distribution and refers to them as popular items. Subsequently, the remaining items found at the long tail of the power law distribution constitute less popular or niche items. 

Given the popular set of items $\mathcal{V}_{\mathrm{pop}}$ and the niche set of items $\mathcal{V}_{\mathrm{niche}}$, then, for each position $t$ in the sequence $\mathcal{S}^{u}$, we classify an item, $v_t$, as popular if $v_t \in \mathcal{V}_{\mathrm{pop}}$ or niche if $v_t \in \mathcal{V}_{\mathrm{niche}}$. The counts of popular and niche items are then aggregated for each position across all sequences. Using this simple setting, we obtained the average ratio of popular to niche items for each position over all the sequences as
\begin{equation}
    \mathrm{Ratio}(t) = \frac{\sum_{u=1}^{\vert \mathcal{U} \vert} \mathbb{1}_{\mathcal{V}_{\mathrm{pop}}}(v_{ut})}{\sum_{u=1}^{\vert \mathcal{U} \vert} \mathbb{1}_{\mathcal{V}_{\mathrm{niche}}}(v_{ut})}
\end{equation}
where $v_{ut}$ is the item that user $u$ interacted with at position $t$. A high ratio indicates that position $t$ predominantly consists of popular items, whereas a low ratio means that position $t$ consists of mostly less popular items.

Fig.~\ref{figure:Ratio} plots this ratio for each position $t$ in each user's sequence of items. It is evident that the higher the position of item $v$, the lower the ratio of popular to niche items. The graph resembles an inversely proportional graph. As the position increases, the number of popular items decreases while the number of niche items increases. This relationship persists throughout all four datasets used.

Given the observed outcomes, the study addresses the shift in preference through a co-attention mechanism to produce a popularity-weighted sequence representation and cross-pseudo supervision to learn from evolving future preferences.

\begin{figure*}[htb!]
\centerline{\includegraphics[width=1.0\textwidth]{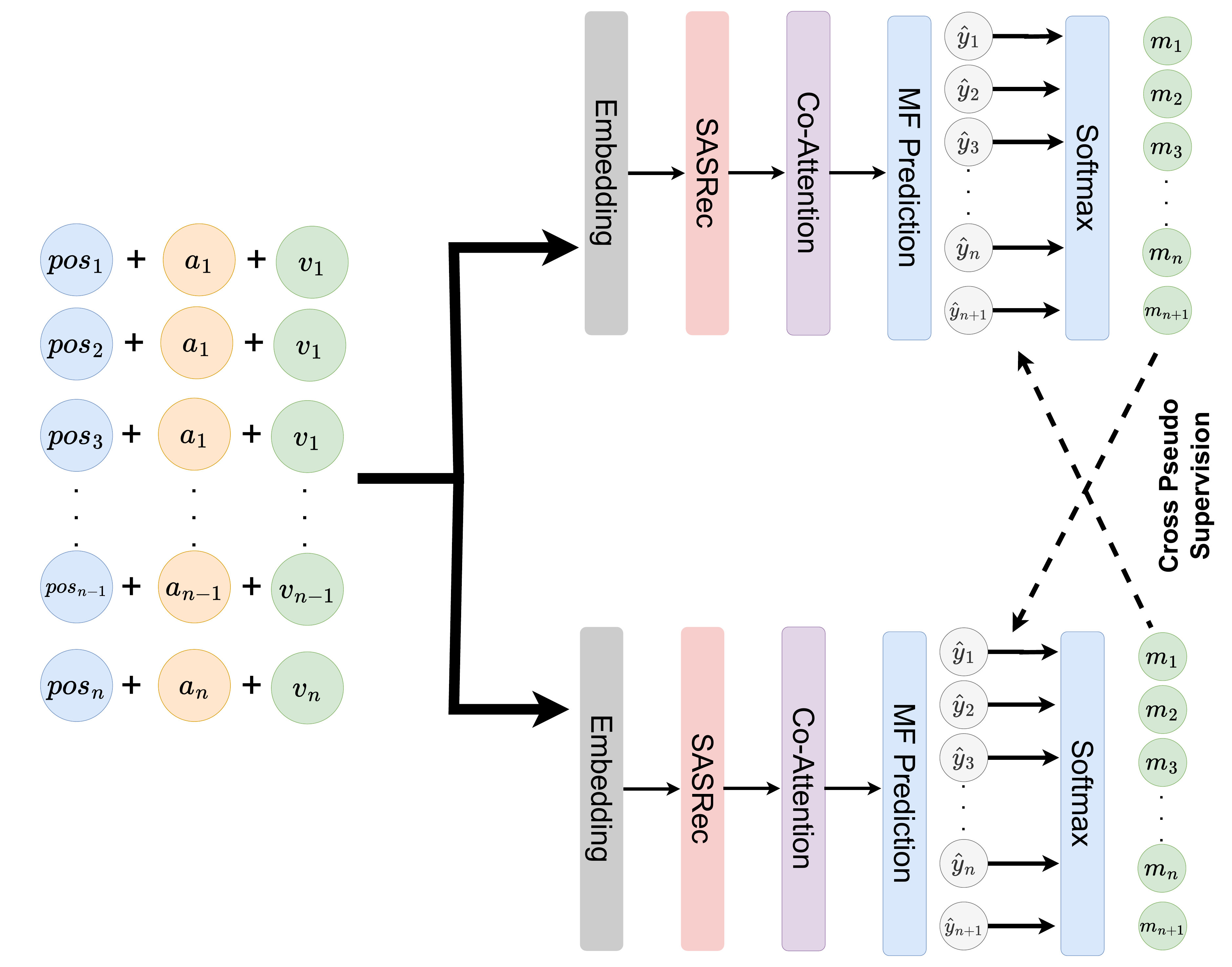}}
\caption{An illustration of the architecture of BiCoRec. The summed positional, auxiliary and item embeddings are passed into four layers to obtain the popularity-weighted sequence representation. Then, cross-pseudo supervision is used to enforce the consistency of predictions between the two networks.}
\label{figure:FinalModel}
\end{figure*}

\section{The Proposed Model}\label{section:Model}

The framework of the proposed \underline{\textbf{Bi}}as-Mitigated \underline{\textbf{Co}}ntext-Aware Sequential \underline{\textbf{Rec}}ommender (BiCoRec) is depicted in Fig.~\ref{figure:FinalModel}. It can be divided into three functional components: the input embedding, attention, and predictive layers. We discuss each component, respectively, in the following subsections. 

\subsection{Embedding Layer}

\subsubsection{Input Embedding}

We produce the learnable embedding $\mathbf{E}_{\mathcal{V}} \in \mathbb{R}^{\vert \mathcal{V} \vert \times d}$ which maps the ID of an item to a vector of dimension $d$. Each position $t$, where $v_{t}^{u} \in \mathcal{S}^{u}$ ($1 \leq t \leq n$), is mapped to a learnable embedding $\mathbf{P} \in \mathbb{R}^{n \times d}$. We consider the text, image, audio and/or tabular data associated with an item as auxiliary information. We obtained the pre-trained vector representations of each modality from a vision, audio or multi-modal model from HuggingFace~\cite{huggingfacedata2vec}. The vector representations from different modalities are concatenated into a single vector $\mathbf{a}_{v} \in \mathcal{A}$ such that $\mathcal{A} = \{\mathbf{a}_1, \mathbf{a}_2, \dots, \mathbf{a}_{\vert \mathcal{V} \vert}\}$ denotes the set of concatenated auxiliary information vectors corresponding to each item. We use a linear transformation to transform the vector into a $ d$-dimensional one, denoted by $\mathbf{e}_{\mathbf{a}_{v}}$.

The final input embedding $\mathbf{E} \in \mathbb{R}^{n \times d}$ is produced by summing the item $\mathbf{E}_{\mathcal{V}}$, position $\mathbf{P}$ and auxiliary embeddings $\mathbf{E}_{\mathcal{A}}$ as
\begin{equation}
    \mathbf{E} =    \begin{bmatrix}
                        \mathbf{e}_{v_{1}^{u}} + \mathbf{p}_{1} + \mathbf{e}_{\mathbf{a}_{v_{1}^{u}}} &
                        \mathbf{e}_{v_{2}^{u}} + \mathbf{p}_{2} + \mathbf{e}_{\mathbf{a}_{v_{2}^{u}}} &
                        \dots &
                        \mathbf{e}_{v_{n}^{u}} + \mathbf{p}_{n} + \mathbf{e}_{\mathbf{a}_{v_{n}^{u}}} \\
                    \end{bmatrix}^\mathrm{T}
\end{equation}
where each vector component corresponds to an item $v_{t}^{u}$ within the sequence $\mathcal{S}^{u}$.

\subsubsection{Popularity Embedding}

First, we quantify the bias inherent within each item. Our method then derives a new sequence representation that describes the popularity of each item in the user's sequence of preferences. Finally, we reduce bias in the latent space by leveraging this popularity embedding to dynamically adjust the attention weights applied to the user’s sequence of preferences. We measure popularity bias by calculating the Term Frequency-Inverse Document Frequency (TF-IDF) for each item. Items unique to a small percentage of sequences (i.e., niche items) receive higher importance than items common across all sequences. Given an item $v_t$ from the user's sequence of items $\mathcal{S}^{u}$, we have the TF-IDF score $q_{ut}$ that describes the item's popularity concerning other users and items. Hence, the vector $\mathbf{q}_{u}$ can describe the user's preference for popular or unpopular items.

The popularity embedding encapsulates each user's preferences for popular items, such that:
\begin{equation}
    \mathbf{E}_{\mathbf{Q}} = \mathbf{q} \mathbf{W} + \mathbf{b}
\end{equation}
where $\mathbf{E}_{\mathbf{Q}} \in \mathbb{R}^{\vert \mathcal{U} \vert \times d}$, $\mathbf{W} \in \mathbb{R}^{d \times d}$ is the weight matrix and $\mathbf{b} \in \mathbb{R}^{d}$ is the bias vector. However, we do not fuse the popularity embedding into the transformer input embedding. Bias is already a characteristic inherent within users and items and should be reflected within their features~\cite{Koren2009Matrix}. The transformer inadvertently absorbs this bias. Therefore, we explicitly incorporate bias mitigation into the user's sequential representation. This approach informs the model about the inherent bias present in each user's sequence.

\subsection{Attention Layer}

\subsubsection{Self-Attention}

Let $\mathbf{H}^{0} = \mathbf{E}$ then we obtain the sequence representation 
\begin{align}
    \mathbf{F}^{l} &= \mathrm{LayerNorm}(\mathbf{H}^{l} + \mathrm{Dropout}(\mathrm{SA}(\mathbf{H}^{l}))) \\
    \mathrm{Trm}(\mathbf{H}^{l}) &= \mathrm{LayerNorm} \left( \mathbf{F}^{l} + \mathrm{Dropout}(\mathrm{PFFN}(\mathbf{F}^{l})) \right) \\
    \mathbf{H}^{l+1} &= \mathrm{Trm}(\mathbf{H}^{l}) 
\end{align}
where $l$ is the number of layers, SA($\bullet$) is the Self-Attention mechanism, PFFN($\bullet$) is the Point-wise FeedForward Network and $\mathrm{Trm}$ is the tranformer~\cite{kang2018self}.

\begin{figure}[htb!]
    \centerline{\includegraphics[width=1.2\textwidth]{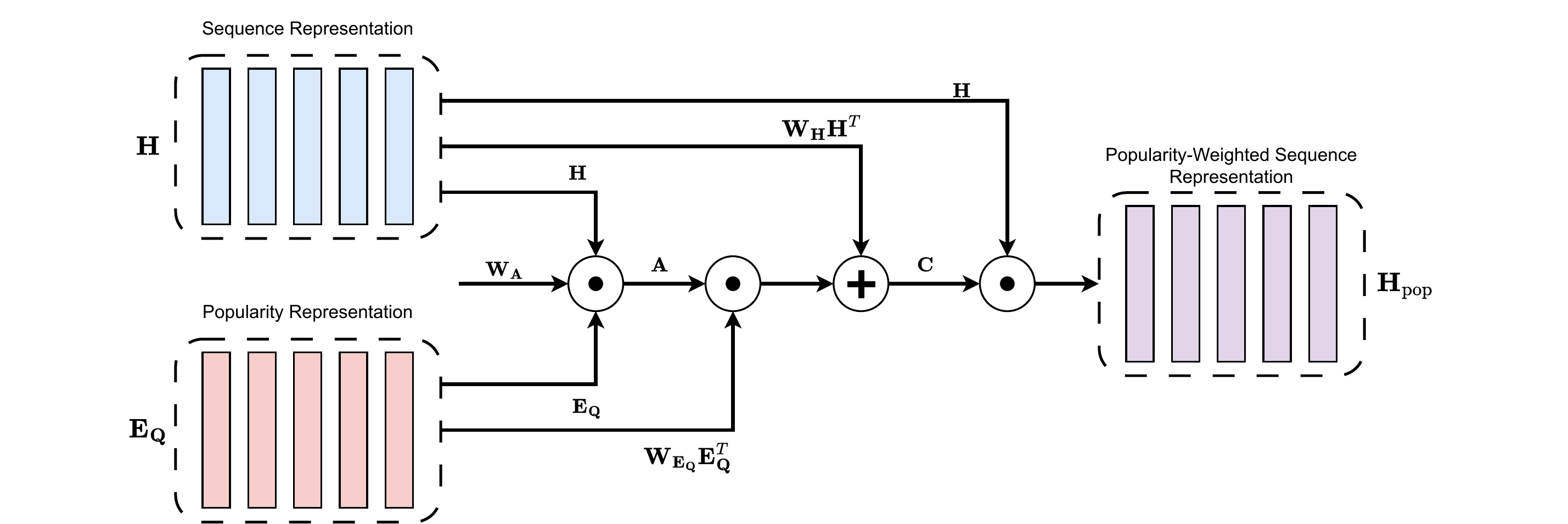}}
    \caption{The parallel co-attention mechanism attends to the sequence and popularity representation simultaneously and then produces a popularity-weighted sequence representation. The $\bigodot$ symbol represents the dot product operation, and $\bigoplus$ refers to the summation operation.}
    \label{figure:CoAttentionMechanism}
\end{figure}

\subsubsection{Co-Attention}

Next, we reduce the bias inherent in the learnt sequence representation $\mathbf{H} \in \mathbb{R}^{n \times d}$ obtained from point-wise FFN. The popularity sequence representation $\mathbf{E}_{\mathbf{Q}}$ describes the latent bias in the user's preferences. We adopted the parallel co-attention mechanism described in Xu and Saenko \textit{et al.}~\cite{xu2016ask}, such that the popularity representation guides sequence attention. Fig.~\ref{figure:CoAttentionMechanism} depicts the diagram of this co-attention mechanism. We compute 
\begin{equation}
    \mathbf{A} = \mathrm{tanh}(\mathbf{H} \mathbf{W}_{\mathbf{A}} \mathbf{E}_{\mathbf{Q}}^{T})
\end{equation}
such that $\mathbf{A} \in \mathbb{R}^{n \times n}$ is the affinity matrix and $\mathbf{W}_{\mathbf{A}} \in \mathbb{R}^{d \times d}$ are the learnable  weights. The affinity matrix is the similarity between the sequence, $\mathbf{H}$, and popularity features, $\mathbf{E}_{\mathbf{Q}}$, at all pairs of sequence and popularity positions. We consider the affinity matrix a feature matrix representing the sequence and popularity attention maps. We then derive an attention map for the sequence representation, $\mathbf{H}$, using the affinity matrix, $\mathbf{A}$. The popularity attention space is first transformed into the sequence attention space to produce the attention map
\begin{equation}
    \mathbf{C}_{\mathrm{map}} = \mathrm{tanh}(\mathbf{W}_{\mathbf{H}} \mathbf{H}^{T} + (\mathbf{W}_{\mathbf{\mathbf{E}_{\mathbf{Q}}}} \mathbf{E}_{\mathbf{Q}}^{T}) \mathbf{A}) 
\end{equation}
$\mathbf{W}_{\mathbf{H}},\mathbf{W}_{\mathbf{E}_{\mathbf{Q}}} \in \mathbb{R}^{k \times d} $ are the weights that transform the sequence $\mathbf{H}$ and popularity $\mathbf{E}_{\mathbf{P}}$ features, $\mathrm{tanh}$ constraints the values from exploding towards infinity and also considers the negative and positive associations in the affinity matrix. $\mathbf{C}_{\mathrm{map}} \in \mathbb{R}^{k \times n}$ highlights parts of the sequence representation that are surpassed by the popularity signal. Now, we compute the attention probabilities as
\begin{equation}
    \mathbf{C}_{\mathrm{prob}} = \mathrm{softmax}(\mathbf{W}_{\mathbf{H}}^{T} \mathbf{C}_{\mathrm{map}})
\end{equation}
and then 
\begin{equation}
    \mathbf{H}_{\mathrm{pop}} = (\tau \mathbf{C}_{\mathrm{prob}}) \mathbf{H}
\end{equation}
which represents the popularity-weighted sequence representation obtained through parallel co-attention, and $\tau$ is the temperature parameter.

\subsection{Prediction Layer}

We used the final sequence representation $\mathbf{H}^{l+1}_{\mathrm{pop}}$ to predict the next item $v_{n+1}$ given the previous $n$ items. We adopt a matrix factorization prediction layer to predict the relevance of item $v_{n+1}$~\cite{kang2018self}: 
\begin{equation}
    \hat{y}_{u v_{n+1}} = \max (\mathbf{h}_{n+1}^{l+1} \mathbf{E}_{\mathcal{V}}^{T})
\end{equation}
where $\hat{y}_{u v_{n+1}}$ is the preference score of item $v_{n+1}$ for the user $u$, $\mathbf{h}_{n+1} \in \mathbb{R}^{d}$ and $\mathbf{E}_{\mathcal{V}}^{T} \in \mathbb{R}^{\vert \mathcal{V} \vert \times d}$. The dot product in MF captures the overall interest in the item’s characteristics, given our knowledge of the user's past interactions. Alternatively, a $\mathrm{softmax}$ layer can also be used; however, MF can efficiently handle a large number of items $\vert \mathcal{V} \vert$. We select the item that generated the highest preference score. 

\begin{figure}[htb!]
\centerline{\includegraphics[width=0.9\textwidth]{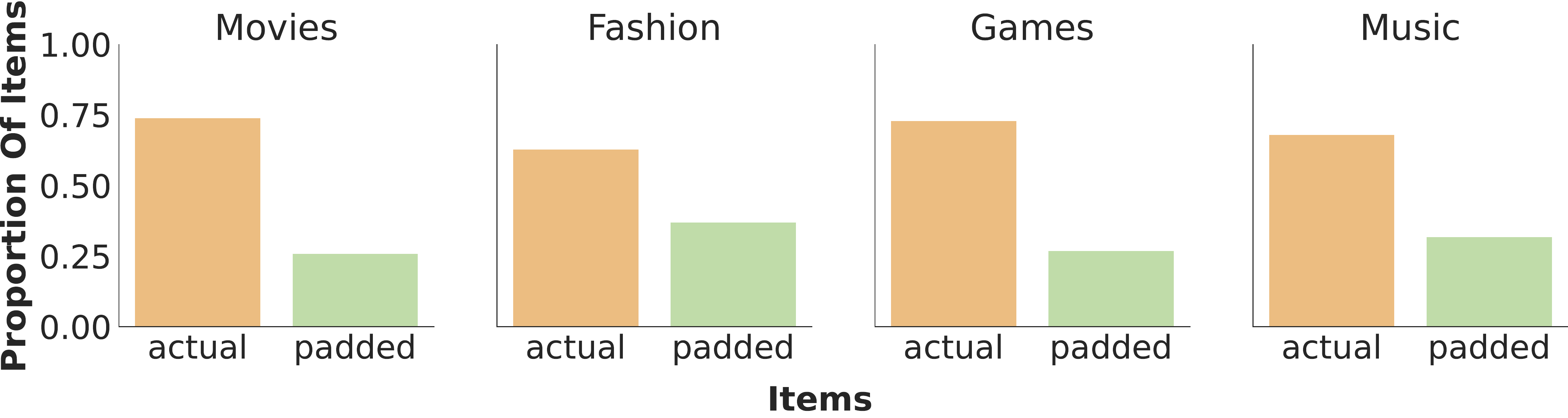}}
\caption{Depicts the proportion of actual and padded items across all the sequences. The padded items are null entries. More than 25\% of the items in a user's sequence are padded items.}
\label{figure:Padded}
\end{figure}

\subsection{Training}

In sequential recommendation, we often define a fixed sequence length $n$ that represents each user's list of interactions. If a user has interacted with more items than $n$, we truncate their sequence of items. Otherwise, if a user has interacted with fewer items than $n$, we pad the sequence with zeros. As a result, many padded items are included in the user's sequence of items. An illustration of the percentage of actual and padded items in each dataset is depicted in Fig.~\ref{figure:Padded}.

The current state-of-the-art models ignore the padded items. For instance, SASRec predicts the representation of the padded items but does not attempt to train on them since their true items are unknown. SASRec only trains on the <next> token. BERT4Rec randomly masks a proportion of all items in the input sequence~\cite{taylor1953cloze}. Given an item, BERT4Rec replaces it with a special token ``[mask]'' and then predicts the original IDs of the masked items based solely on its left and right context. This training process expands the number of samples for training the model by simply masking more items. However, BERT4Rec still ignores the padded items.

More than 25\% of the items are padded items in the user sequences considered, as shown in Fig.~\ref{figure:Padded}. Since a sequence is arranged from oldest to most recent items, the padded items always appear before the actual items. If we ignore the padded items, we fail to train our model on the user's changing future preferences. 

Suppose we train using the <next> token; then, we would only learn how to make short-term predictions. On the other hand, training with the [mask] token involves learning from shorter sequences to make short-term predictions. Instead, we predict the preference score for all items $v \in \mathcal{V}$ for the position $t$ in the users sequence $\mathcal{S}^{u}$ that satisfies
\begin{equation}
    v_{t} = \begin{cases}
                \text{<pad>} & \text{if $v_t$ is a padded item} \\
                \text{<next>} & \text{the next item given our knowledge of previous $t-1$ items }
            \end{cases}
\end{equation}
and then we train for the padded and next items in the sequence. As a result, we train the model using the user's entire sequence of items. This approach is discussed in the following section. 

\begin{figure*}[htb!]
\centerline{\includegraphics[width=1.1\textwidth]{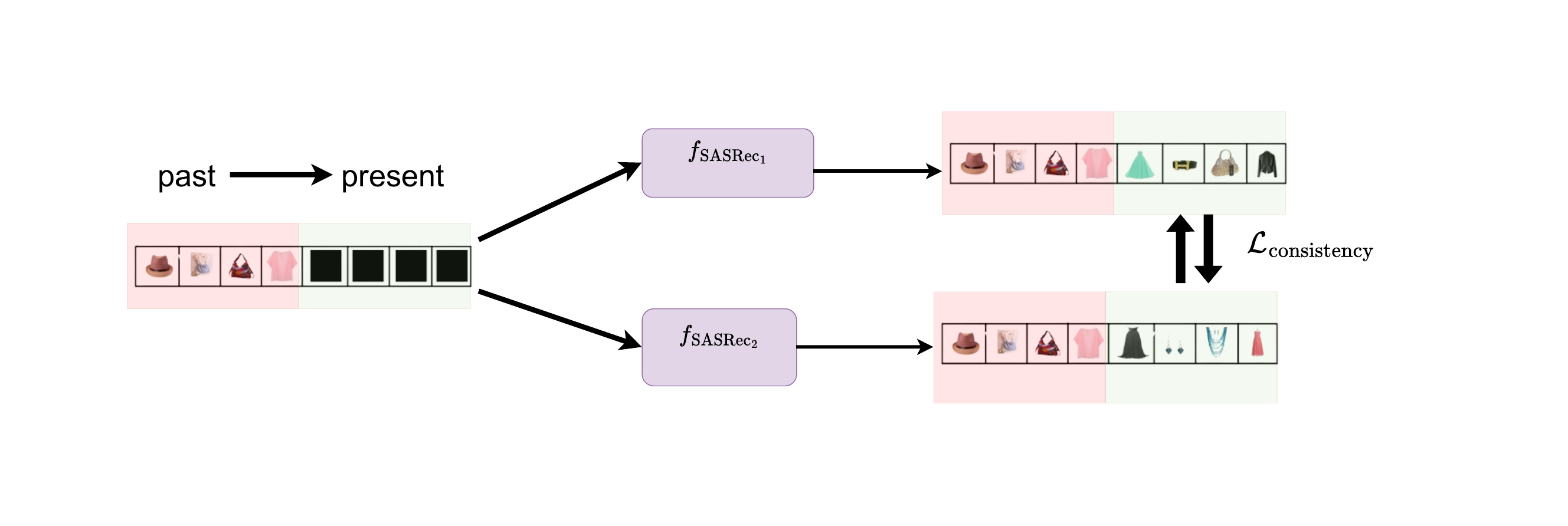}}
\caption{Depicts a user's sequence of items with padded items represented in black. Two SASRec models initialized with different weights predict the padded items (in green). The unsupervised loss function ensures consistency of predictions between the two models.}
\label{figure:CrossPseudo}
\end{figure*}

\subsection{Cross-Pseudo Supervision} 

We may view the actual items as labelled data and the padded items as unlabeled data. The existence of labelled and unlabeled data transforms the problem into a semi-supervised learning paradigm. We adapt the cross-pseudo supervision approach proposed by Chen \textit{et al.}~\cite{chen2021semi} in our sequential recommendation model to train it using labelled and unlabelled data. The main objective of cross-pseudo supervision is to ensure consistency of predictions from both past and future preferences. In our case, the actual items are the previously observed items, and the padded items are the future unobserved items. The model learns a latent representation of the user’s preferences based on the sequence of actual items, while simultaneously generating pseudo-labels for the padded items. Cross-pseudo supervision then ensures that the latent preference profile derived from the padded items aligns with that of the actual items. This process maintains long-term preference consistency by learning from item-to-item transitions. 

The proposed cross-pseudo supervision network used in our sequential recommendations is illustrated in Fig.~\ref{figure:CrossPseudo}. We use a dual network to ensure consistent predictions across differently initialized networks. The objective function then optimizes two loss functions: the supervision loss $\mathcal{L}_{\mathrm{supervised}}$ and the consistency loss $\mathcal{L}_{\mathrm{consistency}}$. Below, we detail the computation of the training loss:
\begin{itemize}
    \item Let $\mathcal{S}_{u}$ be the sequence of items for user $u$ and let $\mathcal{A}$ be the set of auxiliary information vectors corresponding to each item in $\mathcal{S}_{u}$. We pass the input sequence
    \begin{eqnarray}
        (\mathcal{S}_{u}, \mathcal{A}) & \rightarrow f_{\mathrm{SASRec}}((\mathcal{S}_u, \mathcal{A}):\theta_1) \rightarrow \mathbf{\hat{S}}_{u}^{(1)} \rightarrow \mathbf{M}_{u}^{(1)} \\
         & \searrow f_{\mathrm{SASRec}}((\mathcal{S}_u, \mathcal{A}):\theta_2) \rightarrow \mathbf{\hat{S}}_{u}^{(2)} \rightarrow \mathbf{M}_{u}^{(2)}
    \end{eqnarray}
    where two $f_{\mathrm{SASRec}}$ networks maintain the same structure, but their weights, i.e., $\theta_1$ and $\theta_2$, are initialized differently. $\mathbf{\hat{S}}_{u} = \{ \mathbf{\hat{s}}_1, \dots, \mathbf{\hat{s}}_n \text{ | } \mathbf{\hat{s}} \in \mathbb{R}^{\vert \mathcal{V} \vert} \}$ is the pseudo-sequence confidence map obtained from the network and $\mathbf{M}_{u} = \{ \mathbf{m}_1, \dots, \mathbf{m}_n \text{ | } \mathbf{m} \in \{0, 1\}^{\vert \mathcal{V} \vert} \} $ is the predicted one-hot label map.

    \item Supervision loss uses the standard cross-entropy loss $\ell_{ce}$ on the labelled data over the two parallel networks
    \begin{equation}
        \mathcal{L}_{\mathrm{supervised}}^u = \frac{1}{\vert \mathcal{S}^{u} \vert} \sum_{t \in \{ 1, \dots ,n \} } (\ell_{ce}(\mathbf{\hat{s}}_{t}^{(1)}, \mathbf{s}_{t}) + \ell_{ce}(\mathbf{\hat{s}}_{t}^{(2)}, \mathbf{s}_{t}))
    \end{equation}
    where $t$ is the position of the item that satisfies $v_t = \text{<next>}$ and $\mathbf{s}_{t} \in \{0, 1\}^{\vert \mathcal{V} \vert}$ is the ground truth one-hot label that indicates the position of the actual item $v$.

    \item The consistency loss is bidirectional:
    \begin{equation}
        \mathcal{L}_{\mathrm{consistency}}^u = \frac{1}{\vert \mathcal{S}^{u} \vert} \sum_{t \in \{ 1, \dots ,n \} } (\ell_{ce}(\mathbf{\hat{s}}_{t}^{(1)}, \mathbf{m}_{t}^{(2)}) + \ell_{ce}(\mathbf{\hat{y}}_{t}^{(2)}, \mathbf{m}_{t}^{(1)}))
    \end{equation}
    where $t$ is the position of the item that satisfies $v_t = \text{<pad>}$. The one-hot label output $\mathbf{M}_1$ from one network $f(\theta_1)$ is used to supervise the pseudo-sequence confidence map $\mathbf{\hat{S}}_2$ of the other network, and so on.

    \item We jointly optimize the supervision and consistency loss
    \begin{equation}
        \mathcal{L}_{\mathrm{crosspseudo}} = \mathcal{L}_{\mathrm{supervised}} + \lambda \mathcal{L}_{\mathrm{consistency}}
    \end{equation}
    where $\lambda$ is a hyperparameter that controls the contribution from each component. 
\end{itemize}

\begin{table}[htb!]
\centering
\caption{The general statistics of each dataset.}\label{table:DatasetStats}
\small 
\begin{tabularx}{\textwidth}{X*{6}{r}}
\toprule
\textbf{\makecell[bl]{Dataset}} & 
\textbf{\makecell[br]{Movies}} & 
\textbf{\makecell[br]{Fashion}} & 
\textbf{\makecell[br]{Games}} & 
\textbf{\makecell[br]{Music}} \\
\midrule
Number of users             & 6041      & 3719      & 6708      & 9873 \\
Number of items             & 3261      & 13198     & 30933     & 159002 \\
Number of interactions      & 998539    & 24097     & 1217931   & 1143299 \\
Average sequence length        & 165.32    & 6.48      & 181.59    & 115.81 \\
Log user-item ratio             & 0.617     & -1.26     & -1.53     & -2.78 \\
Skewness                    & .9650 & .7100 & .6514 & 1.4294  \\
Sparsity                    & 94.93\% & 99.95\% & 99.41\% & 99.93\%  \\
\bottomrule
\end{tabularx}
\end{table}

\section{Experiments} \label{section:Methodology}
In this section, we present our experimental setup and empirical results. 

\subsection{Experimental Setup}

\textbf{Datasets:} We evaluate the proposed sequential recommendation model on four standard benchmark datasets, namely, MovieLens (Movies)~\cite{harper2015movielens}, Amazon Fashion (Fashion)~\cite{ni2019justifying}, Steam Games (Games)~\cite{kang2018self}, and Music4All-Onion (Music)~\cite{Moscati2022music}. Table~\ref{table:DatasetStats} gives an overview of the characteristics of these datasets. Table~\ref{table:MultiModal} provides a list of modalities for each dataset.

\textbf{Evaluation Setup:} We compare the proposed model against the baselines discussed in Section~\ref{section:Background}. Each baseline falls into one of the following three categories: general, context-aware and sequential recommendation models. 
\begin{enumerate}
    \item General recommendation models only consider the user-item interactions to infer predictions. The following models are frequently used as benchmark models in the literature.
    \begin{itemize}
    \item \textbf{PopRec} - Ranks items according to their popularity. 
    \item \textbf{Bayesian Personalized Ranking (BPR)}~\cite{rendle2012bpr} - Aim to find the correct personalized ranking for all items by optimizing the maximum posterior probability. 
    \item \textbf{Neural Collaborative Filtering (NCF)}~\cite{rendle2010factorizing} - Utilizes a multi-layer neural network architecture for collaborative filtering. 
    \item \textbf{Diffusion Recommender (DiffRec)}~\cite{wang2023diffusion} - Adopts a generative denoising process to infer user interactions.
\end{itemize} 
\item Context-aware models incorporate contextual information into the model. These models either consider some auxiliary or popularity data.
\begin{itemize}
    \item \textbf{De-biased Matrix Factorization (MF)}~\cite{Koren2009Matrix} - Projects both user and item features into the same latent factor space whilst mitigating popularity bias.
    \item \textbf{Factorization Machines (FM)}~\cite{rendle2010factorization} - Uses the user-item interaction matrix with auxiliary information.
    \item \textbf{Disentangling Interest and Conformity with Causal Embedding (DICE)}~\cite{zheng2021disentangling} disentangles representations based on the user's true interests and their conformity to popular trends.
\end{itemize}
\item Sequential recommendation assumes that the sequence of items a user interacts with carries important information about their preferences and future actions. We select the following baseline models, often representing the SOTA in this category.
\begin{itemize}
    \item \textbf{Translation-based Recommendation (TransRec)}~\cite{he2017translation} - TransRec is a first-order Markov Chain model. 
    \item \textbf{Convolutional Sequence Embedding Recommender (Caser)}~\cite{tang2018personalized} - It employs a CNN to learn high-order MCs. 
    \item \textbf{Graph Recurrent Unit for Recommendation (GRU4Rec)}~\cite{jannach2017recurrent} - Uses RNNs to model user action sequences for a session-based recommendation. We treat each user's feedback sequence as a session~\cite{kang2018self}.
    \item \textbf{SASRec}~\cite{kang2018self}  - Employs a self-attention mechanism to weigh the importance of different items in a user's interaction history. 
    \item \textbf{BERT4Rec}~\cite{sun2019bert4rec} - Employs a bidirectional Transformer to consider past and future sequence interactions simultaneously.
    \item \textbf{Frequency Enhanced Hybrid Attention Network for Sequential Recommendation (FEARec)}~\cite{du2023frequency} - Utilize a Fourier Transform to capture high-frequency information, which are items with higher interaction rates in the short term.
\end{itemize}
\end{enumerate}

\textbf{Evaluation Metrics:} We assess our models using three accuracy-based metrics; Recall@N, MRR and NDCG@N~\cite{karypis2001evaluation, jarvelin2002cumulated}. In addition, we consider three distinct fairness metrics: Diversity@N, Novelty@N and Serendipty@N~\cite{smyth2001similarity, castells2021novelty}. We ranked the relevant item against the entire item collection for each user and computed the aggregate score attained for each metric~\cite{krichene2020sampled}.

\textbf{Implementation:} The complete code for BiCoRec and the baselines used are available on GitHub~\footnote{https://github.com/pxpana/BiCoRec}. We implemented each baseline using the popular open-source recommendation library RecBole~\cite{zhao2021recbole}. HuggingFace was used to extract vector representations from pre-trained transformer-based models~\cite{huggingfacedata2vec}. Hyperopt is used for hyperparameter optimization~\cite{bergstra2013hyperopt}. The models are implemented using the Pytorch library. All the models are trained from scratch without any pre-training. All models are trained and evaluated on a PC equipped with an Intel(R) Xeon(R) Gold 5315Y CPU @ 3.20GHz and a single NVIDIA RTX A4000 16GB GPU. 

\textbf{Protocol:} Following previous research, we applied the \textit{leave-one-out} strategy to recommend the next item~\cite{kang2018self, sun2019bert4rec}. For each user interaction sequence, we used the last item as the test data and the item before it as the validation data. The remaining items are used for training. We ranked a relevant item against the entire item collection for each user and computed the aggregate score attained for each metric~\cite{krichene2020sampled}. We conducted ten independent runs of the experiments and performed paired t-tests to determine the significance of our approach compared to the baselines.

\textbf{Hyperparameters:} We report the results of each baseline under its optimal hyperparameter settings. We sample 10\% of the training data for each dataset for hyperparameter optimization. Then, we split the training data again into train and validation sets. The hyperparameters are searched using an exhaustive grid search from HyperOpt. Tuning is performed using the validation set, and performance monitoring ceases after 100 epochs with early stopping after 20 steps of patience. We select the parameters that produced the highest NDCG@10 score. For a fair comparison of our model, we fix the number of attention layers to two with a single head, similar to SASRec. We optimize all the methods with Adam optimizer, and the learning rate is set to 0.001. The selected hyperparameters for our model are presented in Table~\ref{Table:Hyperparameters}. 

\begin{figure*}[htb!]
  \begin{subfigure}{\textwidth}
  \centering
    \includegraphics[width=1.0\linewidth]{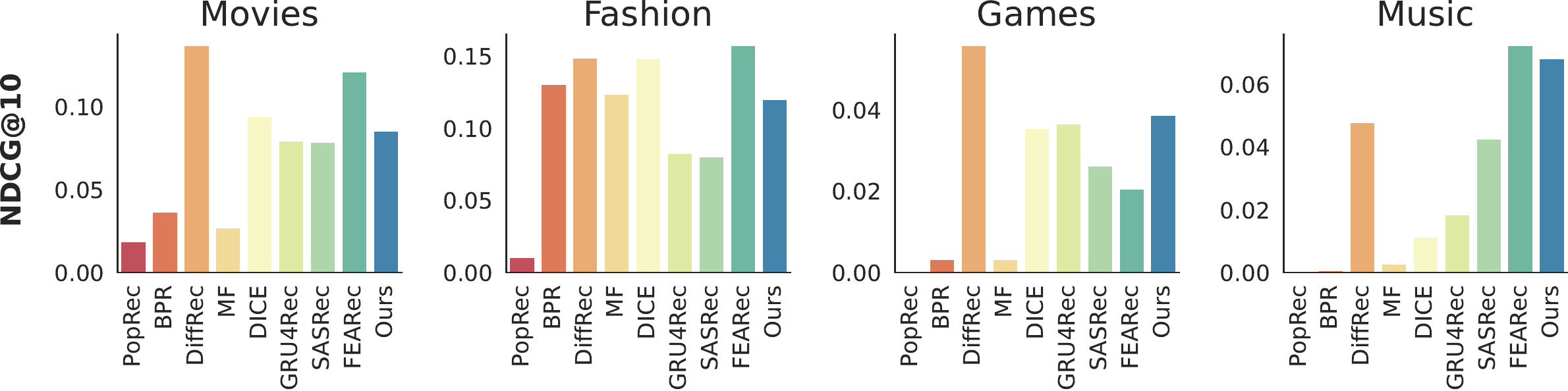}
    \caption{NDCG@10}
    \label{subfig:NDCG_Results}
  \end{subfigure}
  
  \hspace{0.05\textwidth}
  
  \begin{subfigure}{\textwidth}
  \centering
    \includegraphics[width=1.0\linewidth]{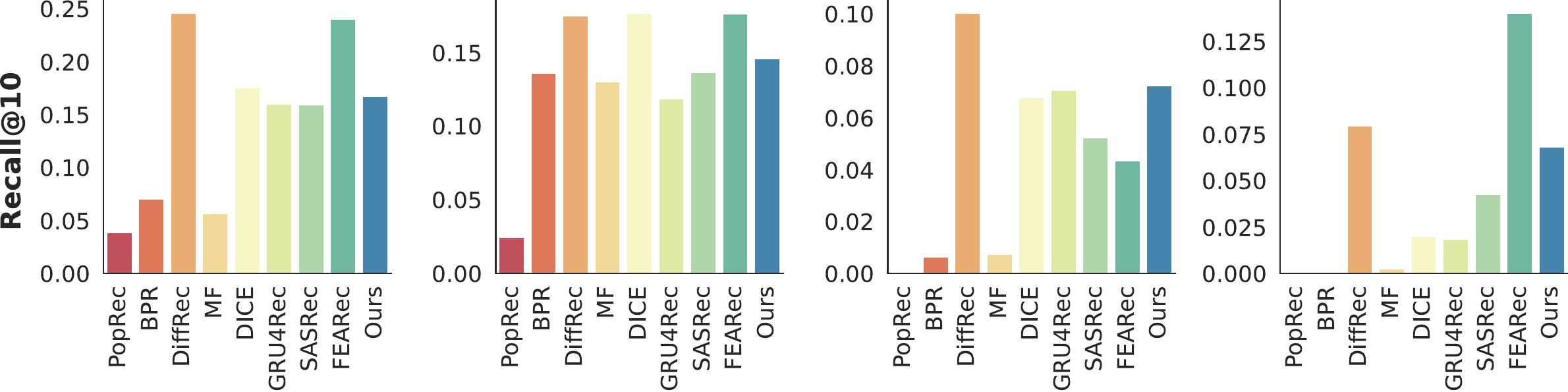}
    \caption{Recall@10}
    \label{subfig:Recall_Results}
  \end{subfigure}

  \caption{The performance of the most consistent models across each dataset. Our proposed approach achieves the highest NDCG@10 and Recall@10 across all datasets except the Fashion dataset. In short-sequence datasets, such as Fashion, biased models like BPR tend to perform better due to the prevalence of popular items within the sequences.}
  \label{figure:Results}
\end{figure*}

\subsection{Results}\label{section:Results}

Fig.~\ref{figure:Results} presents the performance of the most consistent models. The full results are presented in Table~\ref{table:Accuracy}. At first glance, BiCoRec scored the highest against each baseline, except DiffRec and FEARec. BiCoRec struggled to find the relevant item for very short sequences in the Fashion dataset. This performance is expected as short sequences usually consist primarily of popular items, as discussed in Section~\ref{section:ProblemSetting}. 

The recent SOTA models, DiffRec, DICE and FEARec remain superior. BiCoRec achieves a close second or third place for each dataset.

\subsubsection{Popularity Bias}

Next, we categorize users into two groups based on their preferences. We determine a user's preference by observing the last item $v_{n} \in \mathcal{S}^{u}$ in their sequence of preferences. If $v_{n} \in \mathcal{V}_{\mathrm{pop}}$ than the user $u$ prefers a popular item. Conversely, if $v_{n} \in \mathcal{V}_{\mathrm{niche}}$, then the user $u$ prefers a less popular item. Since we now have two sets of users, we calculate the accuracy metrics on the two separate sets. 
\begin{figure*}[htb!]
  \begin{subfigure}{\textwidth}
  \centering
    \includegraphics[width=1.0\linewidth]{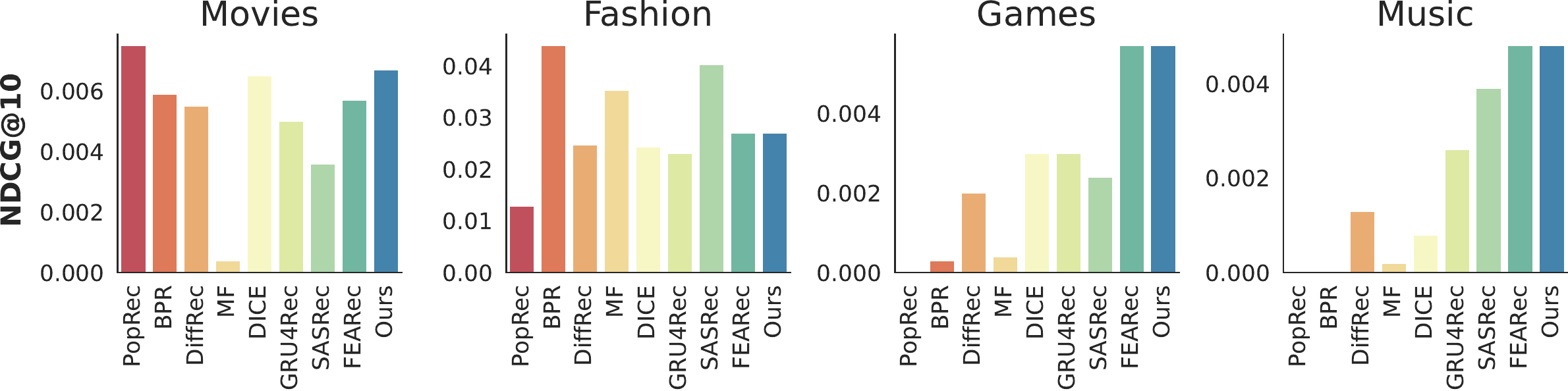}
    \caption{NDCG@10 for users who prefer popular items}
    \label{subfig:ResultsPopularityBias_1}
  \end{subfigure}
  
  \hspace{0.05\textwidth}
  
  \begin{subfigure}{\textwidth}
  \centering
    \includegraphics[width=1.0\linewidth]{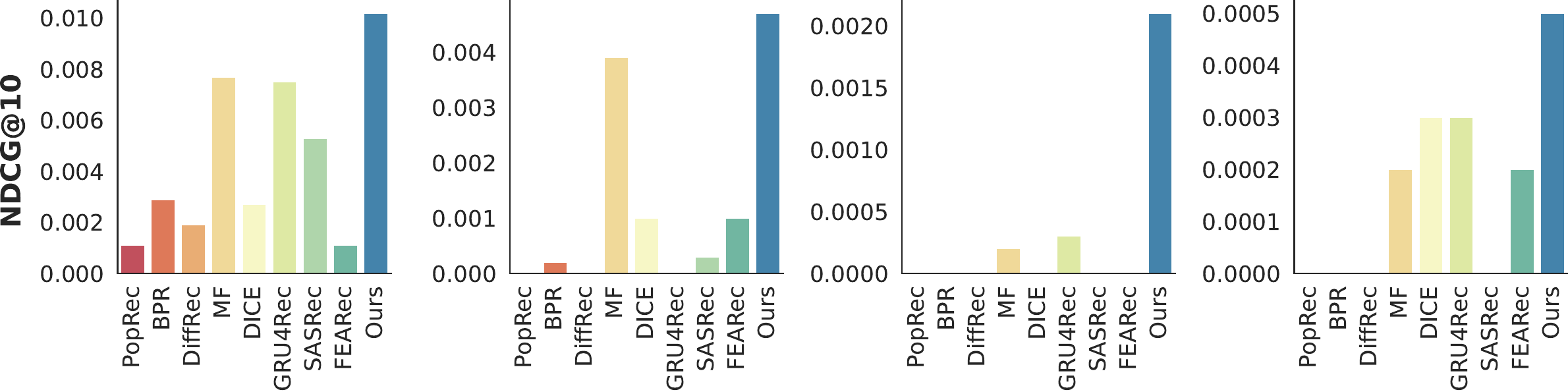}
    \caption{NDCG@10 for users who prefer less popular items}
    \label{subfig:ResultsPopularityBias_2}
  \end{subfigure}

  \caption{The performance over two sets of users with different preferences. BiCoRec achieves the highest NDCG@10 for less popular items. However, our model performs suboptimally over the Movies and Fashion dataset. In contrast, SASRec demonstrates nearly negligible results for less popular items.}
  \label{figure:ResultsPopularityBias}
\end{figure*}

Fig.~\ref{figure:ResultsPopularityBias} provides an overview of the results achieved for users who prefer popular or niche items. Table~\ref{Table:PopularityMetrics} depicts a comprehensive version of the results. The primary contribution to the overall Recall and NDCG score is derived from recommending popular items. Each model's performance in the popular set significantly outweighs that in the less popular set of items. For instance, SASRec scored a Recall@10 of $0.0485$ for users who prefer popular items and performed an order of magnitude less effectively for users inclined towards less popular items with a score of $0.003$.

BiCoRec aimed to mitigate this effect by improving the recommendation performance of users who preferred niche items. For these users, the model achieved improvements of $8.51\%$, $20.51\%$, $23.53\%$, and $25.00\%$ in NDCG@10 over the highest-performing baseline for the Movies, Fashion, Games, and Music datasets, respectively. 

BiCoRec was consistently superior for users who preferred less popular items. The results show a $26.00\%$ average improvement in NDCG@10 over state-of-the-art baselines across all datasets. The primary contribution to the overall Recall and NDCG performance of baseline models is derived from recommending popular items. Consequently, these models optimize themselves
over one group of users. Our bias mitigation technique improved the ranking of items with fewer histories by dynamically re-weighting items in the user’s sequence.

\subsubsection{Change in Preference}

Applying a conventional accuracy metric to all items within a sequence fails to capture the nuanced shift in preference. In short sequences, popularity bias is more prominent, as these sequences often comprise predominantly popular items. Consequently, a model heavily biased toward popular items may perform well in such scenarios. However, its performance diminishes when deploying the same model in an offline setting with considerably longer user sequences.

We employed a sliding window approach for each sequence $\mathcal{S}^{u}$ to evaluate it. Let the length of the sliding window be N=50, then we took a subset of items $\mathcal{S}^{u}_{1:N} = \{ v_{1}, \dots, v_{N} \}$. We predict the next item $v_{N+1}$ given $\mathcal{S}^{u}_{1:N}$. Then we slide the window to get $\mathcal{S}^{u}_{N:N+50}$ and predict $v_{N+50+1}$, and so on. We iteratively perform this process to obtain the NDCG@10 for each window over all the users. 

\begin{figure*}[htb!]
    \centerline{\includegraphics[width=1.0\textwidth]{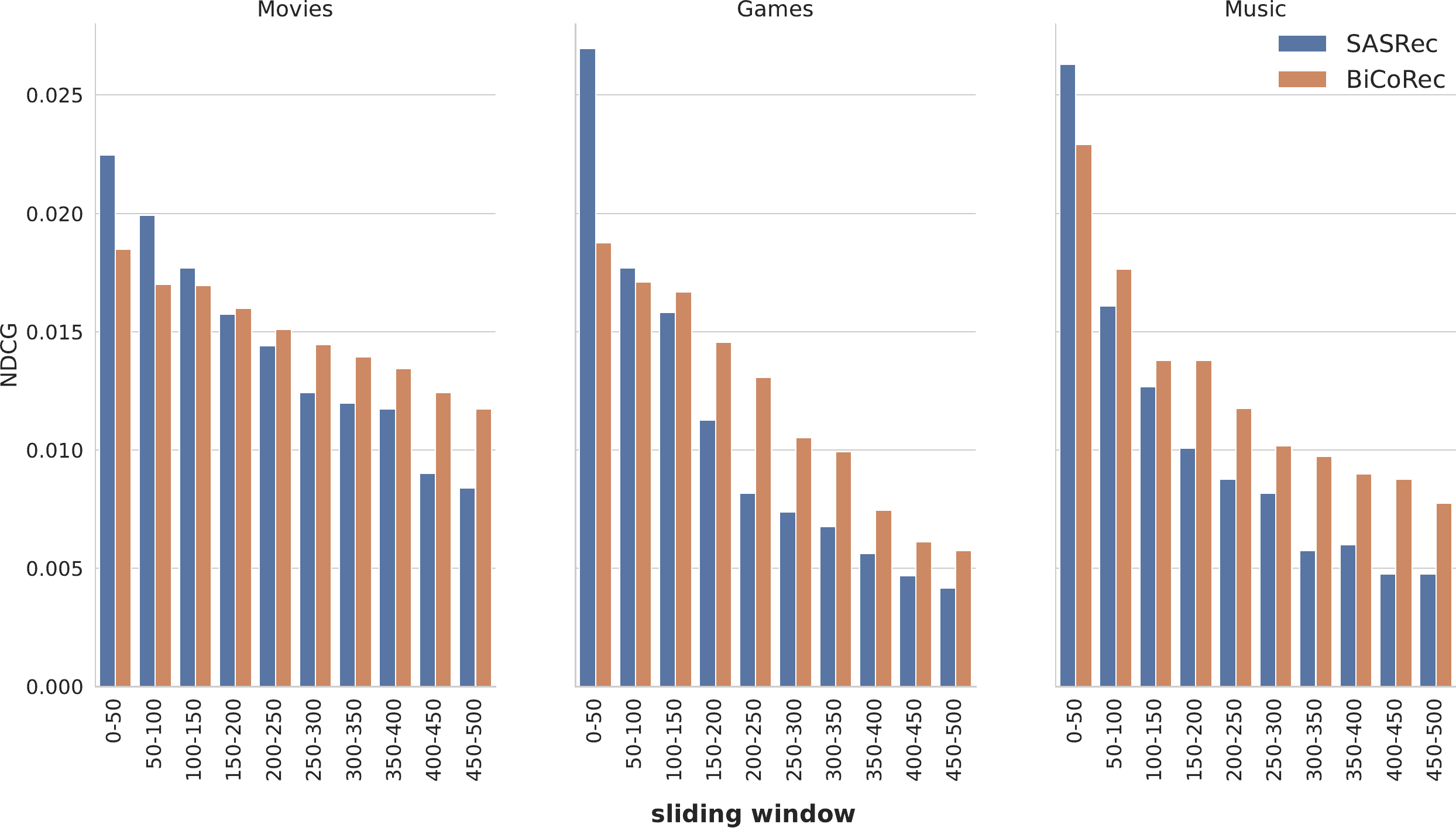}}
    \caption{The performance of SASRec and BiCoRec along the users' sequences through a sliding window of size 50. BiCoRec outperforms SASRec as the window slides across the sequence of items from oldest to recent.}
    \label{figure:SlidingWindow}
\end{figure*}

Fig.~\ref{figure:SlidingWindow} depicts the performance of SASRec and BiCoRec.  SASRec reflects a diminishing NDCG@10 as the window is slid across the sequence. SASRec struggles to adapt to changing preferences over time. The highest NDCG@10 is achieved within the first 50 items in the sequence. The high scores occur when users first start engaging with the system. Hence, they primarily interact with more popular items. However, as we slide the window across to the next 50 items, the NDCG score decreases gradually. The lower scores occur when the users select more niche items that cater to their specific preferences.

BiCoRec also depicts a diminishing NDCG@10 as we slide the window across the sequence. However, we initially obtained a lower NDCG@10 than SASRec around the first 50 to 150 items. As the sequence becomes longer, the overall NDCG@10 obtained is higher than that of SASRec. 

\subsubsection{Precision-Recall Curves}

\begin{figure*}[htb!]
    \centerline{\includegraphics[width=1.0\textwidth]{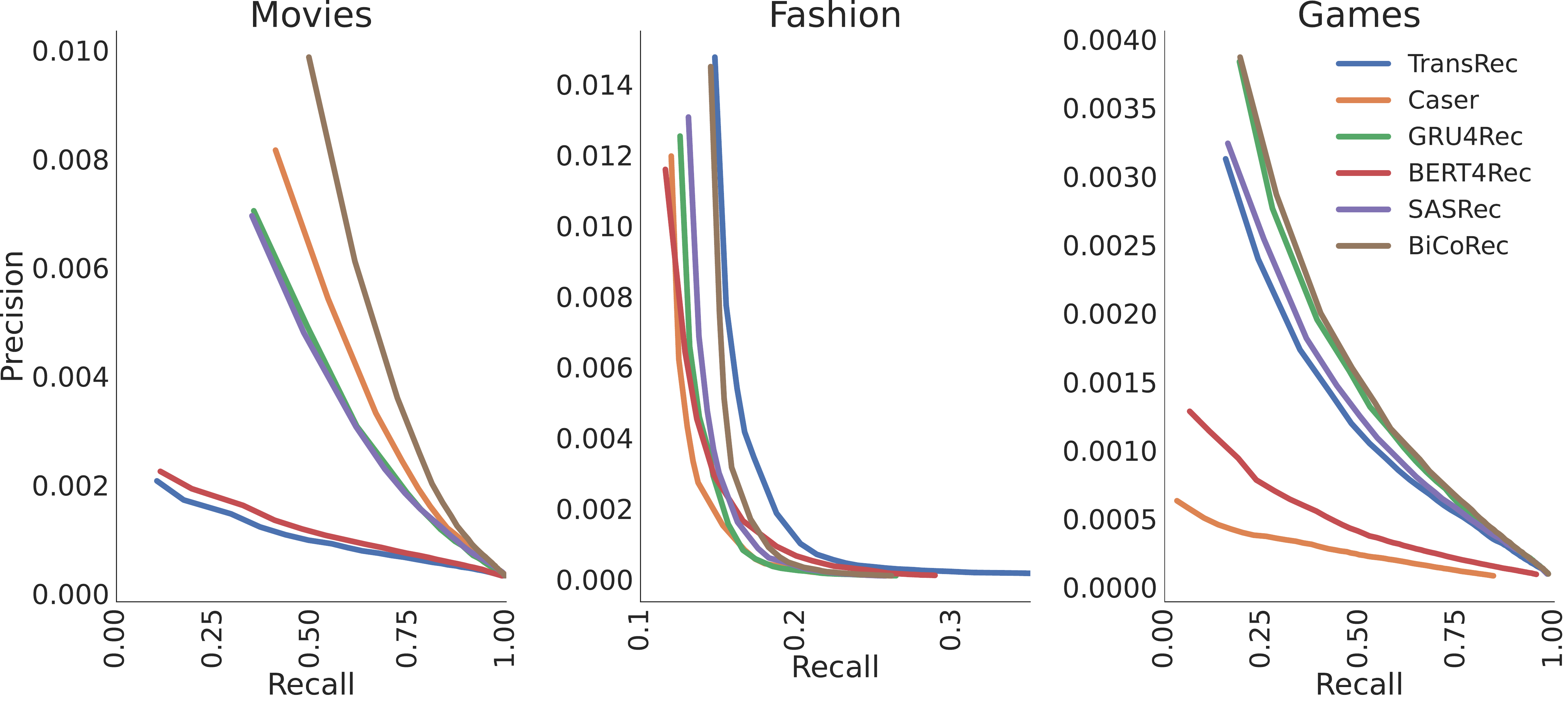}}
    \caption{The precision-recall curves of the SOTA sequential recommendation models and BiCoRec. BiCoRec depicts the steepest slope for long sequence datasets. It falls short of one Fashion dataset model due to its very short sequences.}
    \label{figure:PrecisionRecall}
\end{figure*}

Fig.~\ref{figure:PrecisionRecall} plots the precision-recall curve of each state-of-the-art baseline model and BiCoRec. The curves focus on the trade-off between precision and recall. In particular, the precision-recall curve evaluates the models' ability to detect rare events. Detecting rare events is particularly helpful in the recommendation setting, as the relevant item is much rarer when ranked against the entire collection. At first glance, the precision-recall curves portray an inverse relationship. As recall increases, the precision scores decrease. This trend reveals a moderately good classifier. A more effective setting would be a curve that maintains a high precision score even as recall decreases. BiCoRec depicts the steepest slope for the Movies and Games dataset and is second to TransRec for the Fashion dataset. The Fashion dataset is more skewed toward popular items. Thus, TransRec demonstrates strong retrieval capabilities on highly biased datasets. 

\begin{figure}[htb!]
  \centering
  \begin{tabular}{c}
        \includegraphics[width=1.0\textwidth]{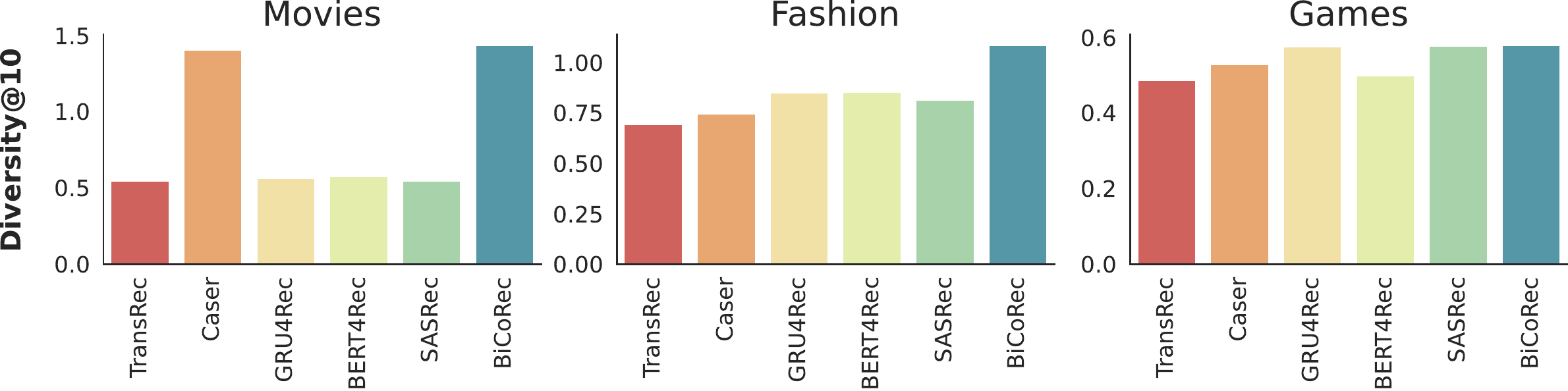} \\ 
        \includegraphics[width=1.0\textwidth]{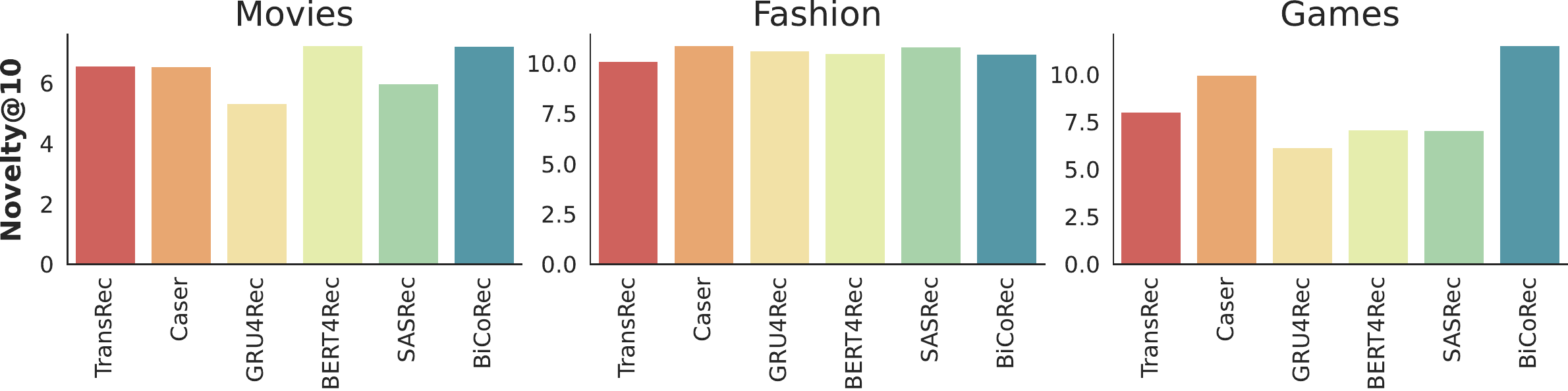} \\
        \includegraphics[width=1.0\textwidth]{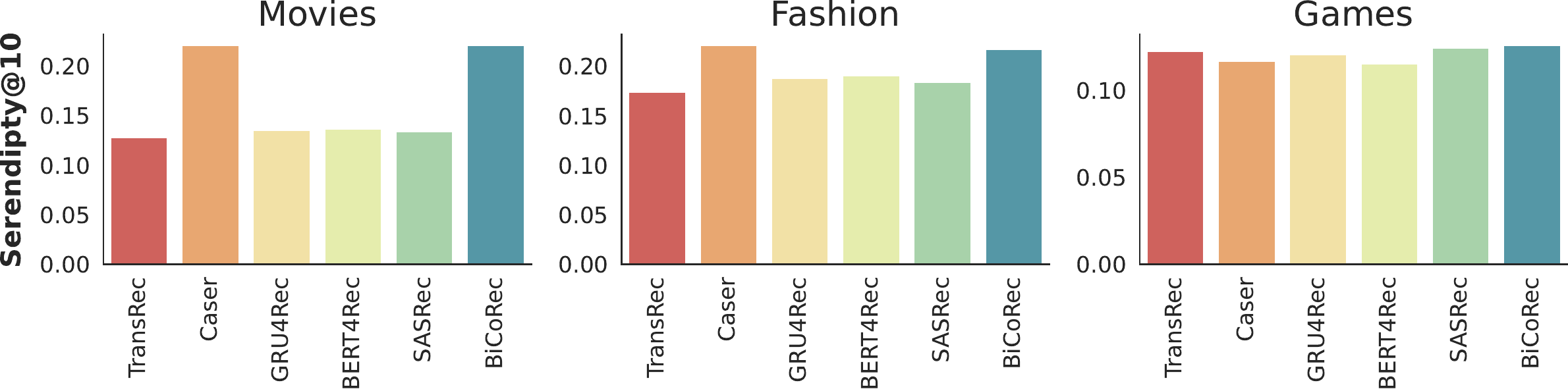} \\ 
  \end{tabular} 
  \caption{The diversity, novelty and serendipity scores achieved by SOTA models and BiCoRec on three datasets.}
  \label{figure:Fairness}
\end{figure}

\subsubsection{Fairness Metrics}

Fig.~\ref{figure:Fairness} presents the diversity, novelty and serendipity scores obtained by each state-of-the-art model. BiCoRec consistently produces a high diversity, serendipity and novelty score. Although the current state-of-the-art model, SASRec, also performs well in recommending novel items, BiCoRec distinguishes itself by prioritizing the relevance of these novel items.

Only Caser achieves near-competitive performance to BiCoRec. Moreover, it often performs second to BiCoRec on accuracy metrics for the Movies, Games and Music datasets. BiCoRec achieved the highest diversity@10 score for the Fashion dataset despite ranking second to TransRec in accuracy metrics. TransRec’s low diversity and novelty scores indicate that it primarily recommends popular items to boost its accuracy on biased datasets. BERT4Rec's high novelty score and low performance on accuracy metrics depict an almost random approach to recommendations.

\begin{table}[htb!]
\caption{The performance of BiCoRec after removing each proposed component.}\label{table:Ablation}
\small
\centering
\begin{tabularx}{\textwidth}{X *{8}{r}}
\toprule
\textbf{Dataset} & \multicolumn{3}{c}{\textbf{Movies}} & & \multicolumn{3}{c}{\textbf{Fashion}} \\

\cmidrule(lr){2-4} \cmidrule(lr){6-8} 

\quad \textbf{Metrics} & \textbf{Recall} && \textbf{NDCG} &
                 & \textbf{Recall} && \textbf{NDCG} \\
\midrule
    \textbf{BiCoRec}                           & \textbf{.1677} && \textbf{.0853} && \textbf{.1459} && \textbf{.1200} \\
    \cmidrule(lr){2-8}
        \quad w/o cross-pseudo              & .1350 && .0535 && .0982 && .0813 \\
        \quad w/o auxiliary information     & .1529 && .0611 && .1170 && .0838 \\
        \quad w/o co-attention layer        & .1498 && .0599 && .1251 && .1112 \\
        \quad w/o user embedding            & .1658 && .0844 && .1438 && .1195 \\

\bottomrule
\end{tabularx}
\end{table}

\subsubsection{Ablation Study}

We performed an ablation study of BiCoRec over two datasets. The critical difference between the Fashion and Movies dataset is that they consist of short and long sequences. Table~\ref{table:Ablation} presents the results of the ablation study. A high recall value means most users have seen their preferred item amongst the $N$ recommended items. A high NDCG suggests that the model frequently ranks the relevant item higher in the list. We tested how removing each proposed component affects the model's overall performance. 

\textbf{Cross-Pseudo Supervision:} Cross-pseudo supervision forces the model to generate consistent representations for past and future items. Removing cross-pseudo supervision in Table~\ref{table:Ablation} led to a $37\%$ and $32\%$ reduction in the NDCG@10 score for the Movies and Fashion datasets, respectively. This is the highest drop in performance. Without the pseudo labels, the representations are learned by predicting the next item given the previous items. Hence, the future items are ignored. 

\textbf{Co-Attention Layer:} The co-attention layer informs the model about the inherent bias present in each user's sequence. Removing the co-attention layer in Table~\ref{table:Ablation} resulted in a $30\%$ and $7\%$ decrease in the NDCG@10 score for the Movies and Fashion datasets, respectively. The Movies dataset has a much longer average sequence length than the Fashion dataset. Hence, user preferences are shifting towards more niche items. The co-attention layer can capture this shift in preferences for longer sequences.

\textbf{Multi-modal Auxiliary Information:} Multi-modal auxiliary information helps improve each item's latent description. Removing auxiliary information in Table~\ref{table:Ablation} resulted in a $28\%$ and $30\%$ reduction in the NDCG@10 score for the Movies and Fashion datasets, respectively. Without this information, the model relies solely on user interaction data. 

\begin{table}[htb!]
\caption{Paired t-test of the NDCG@10 scores achieved by the state-of-the-art sequential recommendation models over BiCoRec's achieved results. The degrees of freedom is fixed at 16}\label{table:Significance}
\centering
\setlength{\tabcolsep}{1.0pt}
\begin{tabular}{llrrrrrrr}
\textbf{\makecell[bl]{Dataset}} & 
&
\makecell[br]{TransRec}    & 
\makecell[br]{Caser}       & 
\makecell[br]{GRU4Rec}     & 
\makecell[br]{BERT4Rec}    & 
\makecell[br]{SASRec}      &
\textbf{\makecell[br]{BiCoRec}}     \\
\midrule
& mean                          & .1127 & .0172     & .0826     & .0485     & .0801     & .1200 \\
\textbf{Fashion} & Standard deviation            & .0098 & .0131     & .0090     & .0051     & .0083     & .0002\\
& t statistic                   & 2.600 & 11.64     & 10.66     & 42.90     & 11.80     & \\
& p-value                       & .0193 & 1.6e-09  & 1.1e-08   & 6.0e-18   & 2.6e-09   & \\
    \cmidrule{2-8}
& mean                          & .0098     & .0810 & .0794 & .0132     & .0786 & .0853 \\
\textbf{Movies} & Standard deviation            & .0004     & .0259 & .0042 & .0029     & .0067 & .0113 \\
& t statistic                   & 19.06     & .5032 & 1.552 & 17.57     & 2.065 & \\
& p-value                       & 2.0e-12   & \textbf{.6217} & \textbf{.1400} & 7.0e-12   & \textbf{.0555} & \\
    \cmidrule{2-8}
& mean                          & .0038     & .0000 & .0366 & .0003     & .0263 & .0387 \\
\textbf{Games} & Standard deviation            & .0010     & 1.5e-05 & .0023 & .0001     & .0088 & .0051 \\
& t statistic                   & 18.99     & 21.44 & 1.107 & 21.40     & 11.76 & \\
& p-value                       & 2.1e-12   & 3.3e-13 & \textbf{.2846} & 3.4e-13   & 2.7e-09 & \\
    \cmidrule{2-8}
& mean                          & .0001     & .0000 & .0083 & .0000     & .0251 & .0352 \\
\textbf{Music} & Standard deviation            & 2.1e-12   & 8.9e-06 & .0005 & 2.7e-05     & .0077 & .0042 \\
& t statistic                   & 23.39     & 23.44 & 17.86 & 23.43     & 3.266 & \\
& p-value                       & 8.5e-14     & 8.1e-14 & 5.5e-12 & 8.2e-14     & .0048 & \\
\bottomrule
\end{tabular} 
\end{table}

\subsubsection{Statistical Significance}

Table~\ref{table:Significance} reports the statistical significance of BiCoRec using the NDCG@10 score. We reject the null hypothesis at a p-value of 0.05; that is, the observed performance of BiCoRec over the baseline model could not have occurred by chance. BiCoRec achieves statistically significant results across all datasets except in four cases (bold-faced in Table~\ref{table:Significance}). These cases occur for Caser, GRU4Rec, and SASRec in the Movies dataset, and for GRU4Rec in the Games dataset. These high p-values stem from BiCoRec's marginal NDCG@10 improvement (see Fig.~\ref{figure:Results} ) over these models.

On average, BiCoRec surpasses baselines, suggesting a real effect, though larger samples may yield more conclusive results. Nonetheless, BiCoRec consistently delivers statistically significant results for the Fashion and Music datasets, excelling in shorter and moderately long sequences.

\section{Conclusion} \label{section:Conclusion}

Current debiasing methods used in traditional recommendation models cannot be applied directly to sequential data due to complicated temporal dependencies. We developed a \textbf{Bi}as-Mitigated \textbf{Co}ntext-Aware Sequential \textbf{Rec}ommendation Model (BiCoRec), designed to adapt to evolving user preferences. A co-attention mechanism simultaneously attends to the user sequence and popularity representations. Our training strategy employs cross-pseudo supervision to learn from evolving user preferences.

On average, BiCoRec is $7\%$ more effective at identifying the most relevant items. Our model achieved a mean increase of $3.14\%$ in NDCG scores on long user sequence datasets, surpassing current state-of-the-art models. Additionally, we observed an average $26\%$ improvement in NDCG for users with niche preferences compared to state-of-the-art baselines across all datasets.

\subsection{Limitation}

BiCoRec may underperform for shorter sequences where users strongly favour popular items. In these cases, popularity-weighted sequences leverage TF-IDF scores to emphasize infrequent items over frequent ones. BiCoRec also showed limited statistical significance in datasets with a large number of items. In such settings, all models tend to yield marginal gains due to the challenge of ranking relevant items within an extensive collection.

\subsection{Future Work}

Future work could explore biases beyond popularity, utilising a more general multi-attention mechanism to dynamically accommodate multiple bias types. Pseudo-labelling requires adequate labelled data to maintain prediction consistency, but due to the high sparsity of recommendation datasets, pseudo-labels may not be ideal. Other semi- or self-supervised methods should be investigated to address these limitations.

\section{Declarations}

\subsection{Data Availability}

The code and weights of the trained models are available here \href{https://github.com/pxpana/BiCoRec}{https://github.com/pxpana/BiCoRec}. Each benchmark dataset is made publicly available by the authors. Specifically we use the MovieLens~\cite{harper2015movielens}, Amazon Fashion~\cite{ni2019justifying}, Steam Games~\cite{kang2018self} and  Music4All-Onion~\cite{Moscati2022music}.

\subsection{Acknowledgements}




\subsection{Conflict of Interest}

The authors declare that there are no conflicts of interest regarding the publication of this paper. The authors confirm that the research was conducted in the absence of any commercial or financial relationships that could be construed as a potential conflict of interest.

\newpage

\bibliography{main}

\newpage

\begin{appendices}

\section{Figures}

\begin{figure*}[htb!]
\centerline{\includegraphics[width=1.0\textwidth]{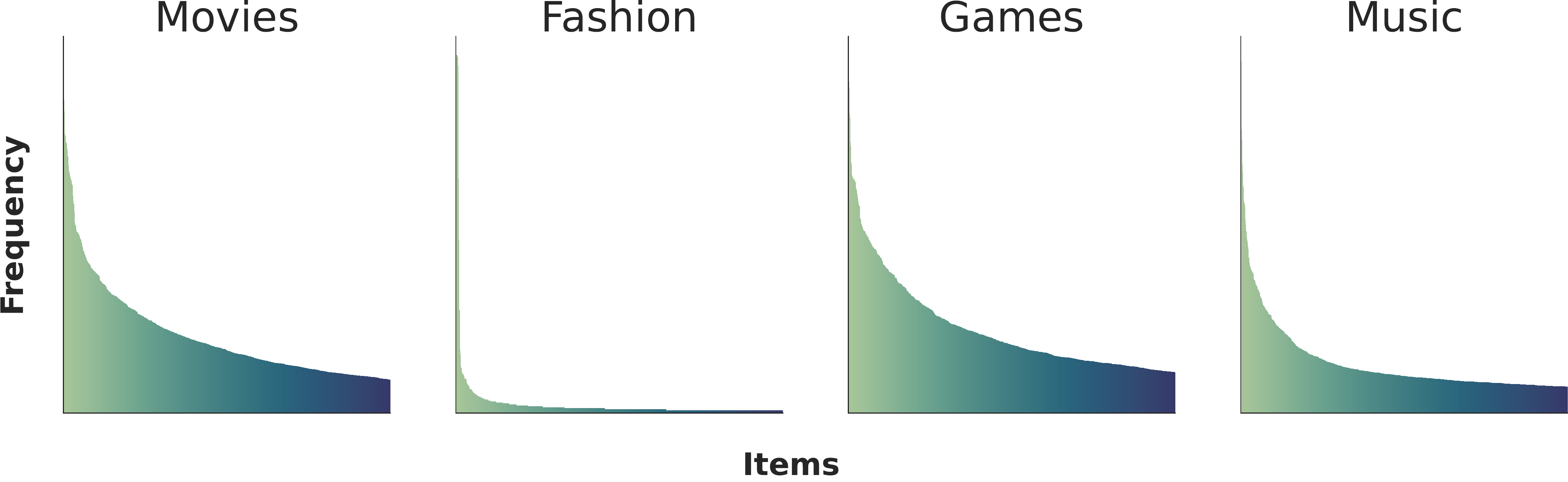}}
\caption{Depicts the frequency of each item in a dataset ordered by its popularity. Each dataset follows a power law distribution. A smaller fraction of items appear more frequently than the majority of items.}
\label{figure:PowerLaw}
\end{figure*}

\begin{figure}[htb!]
    \centerline{\includegraphics[width=1.0\textwidth]{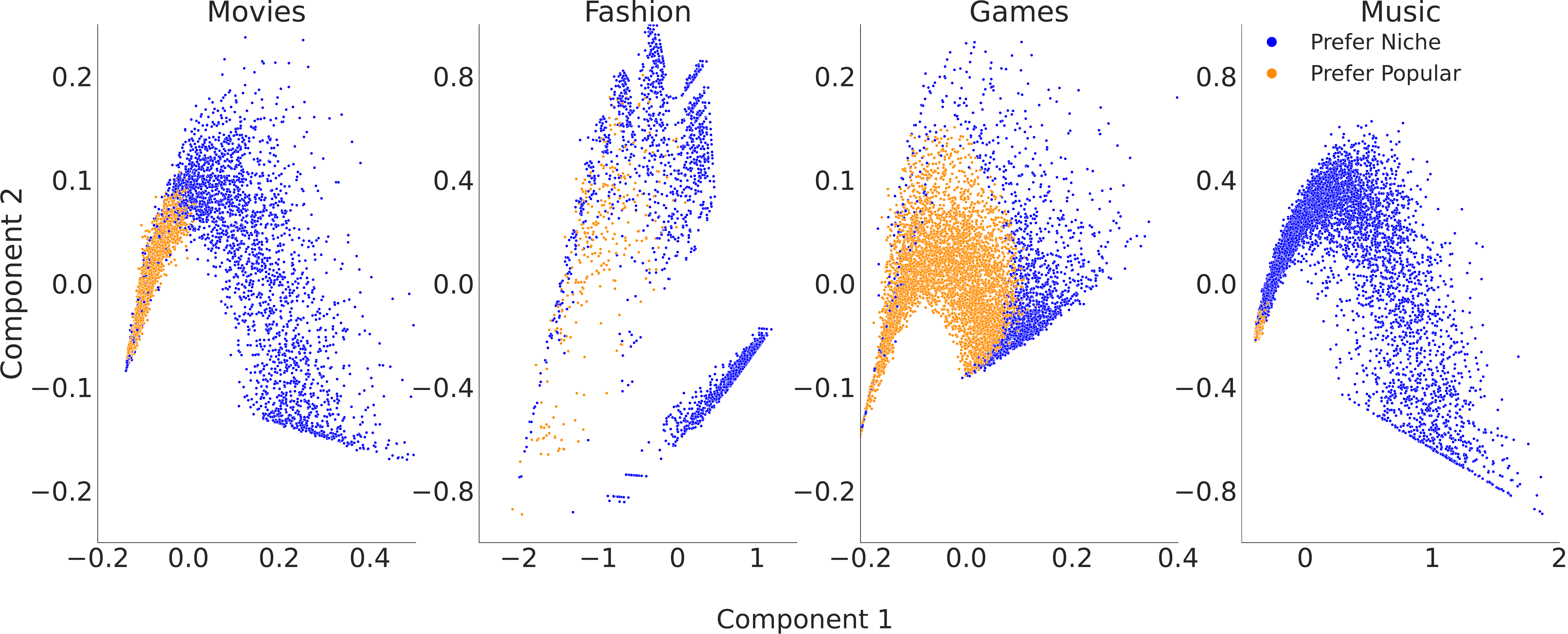}}
    \caption{Two-component PCA projection of each user's sequence of TF-IDF popularity scores. From these scores, we can easily identify each user's preference for popular or unpopular items}
    \label{figure:TFIDF}
\end{figure}

\section{Descriptions}

\subsection{TF-IDF}

The Term Frequency - Inverse Document Frequency (TF-IDF) is calculated for each item. In this setting, the term refers to the item, the document is a sequence of items, and the corpus encompasses all the users' sequences. The Term Frequency (TF) measures the presence of an item within a sequence.
\begin{equation}
    \mathrm{TF}_{v} = \frac{\text{number of times the item $v$ appears in the sequence}}{\text{total number of items in the sequence}}
\end{equation}
which produces $\frac{1}{n_{u}}$ if the item is in the sequence, since each item can appear only once within a sequence. Inverse Document Frequency (IDF) of an item reflects the proportion of corpus sequences containing the item. Items unique to a small percentage of sequences (e.g., niche items) receive higher importance than items common across all sequences.
\begin{equation}
    \mathrm{IDF}_{v} = log \left( \frac{\text{total number of sequences in the corpus}}{\text{number of sequences that contain the item $v$}} \right)
\end{equation}
TF-IDF balances the item commonality within a sequence, measured by TF, with the rarity between sequences measured by IDF, such that: 
\begin{equation}
    \text{TF-IDF}_{v} = \mathrm{TF}_{v} \times \mathrm{IDF}_{v}
\end{equation}
the score reflects the importance of an item for a sequence in the corpus. For each user, we obtain the TF-IDF scores of each item in the sequence and refer to the new sequences as the popularity scores $\mathbf{P} \in \mathbb{R}^{\vert \mathcal{U} \vert \times n}$. That is, for an item $v_t$ from the user's sequence of items $\mathcal{S}^{u}$, we have the TF-IDF score $\mathbf{p}_{ut}$, which describes the item's popularity relative to other users. Hence, the vector $\mathbf{p}_{u}$ describes the user's preference for popular and niche items. 

\subsection{PCA}

We applied PCA to test the effectiveness of our TF-IDF popularity scores. Figure~\ref{figure:TFIDF} depicts a  2-component visualization of the scores. We grouped the users based on their preference for popular and less popular items. The TF-IDF popularity scores can effectively distinguish between users who prefer popular items and those who prefer less popular items. Interestingly, more users prefer less popular items for each dataset.

\section{Tables}

\begin{table}[htb!]
\caption{Different types of auxiliary information for each dataset.}
\centering
\begin{tabularx}{\textwidth}{X *{6}{r}}
\toprule
\textbf{Dataset} & \textbf{Tabular} & \textbf{Text} & \textbf{Image} & \textbf{Audio} \\
\midrule
Fashion & \checkmark & \checkmark & \checkmark & \ding{55} \\
Movies & \checkmark & \checkmark & \checkmark & \ding{55} \\
Games & \checkmark & \checkmark & \ding{55} & \ding{55} \\
Music & \ding{55} & \checkmark & \ding{55} & \checkmark \\
\bottomrule
\end{tabularx}
\label{table:MultiModal}
\end{table}

\begin{table}[!htb]
  \centering
  \caption{Hyperparameters for BiCoRec selected using hyperopt}
  \small
    \begin{tabularx}{\textwidth}{X *{7}{r}}
    \toprule
    \makecell[lb]{
    \textbf{Dataset}} &
    \makecell[rb]{Movies} & 
    \makecell[rb]{Fashion} & 
    \makecell[rb]{Games} &
    \makecell[rb]{Music}  \\
    \midrule
    number of layers          & 2     & 2     & 2     & 2 \\
    number of heads           & 1     & 1     & 1     & 1 \\
    unsupervised loss weight  & 0.3   & 0.1   & 0.4   & 0.3 \\
    co-attention temperature  & 1     & 0.8   & 1     & 1 \\
    layer norm epsilon        & 1e-12 & 1e-6  & 1e-12 & 1e-12 \\
    mask ratio                & 0.6   & 0.4   & 0.3     & 0.7 \\
    gradient clipping         & 5     & 5     & 5     & 10 \\
    weight decay              & 0.001 & 0.0001 & 0.0001  & 0.001 \\
    hidden size               & 64     & 64   & 64     & 128 \\
    hidden dropout            & 0.5     & 0.2     & 0.6     & 0.3 \\
    attention dropout         & 0.2     & 0.4     & 0.2     & 0.4 \\
    \bottomrule
    \end{tabularx}
  \label{Table:Hyperparameters}
\end{table}

\begin{table}[htb!]
\caption{Accuracy Metrics of BiCoRec compared with 12 baselines. The best-performing models are in bold, and the second-best-performing models are underlined.}\label{table:Accuracy}
\small
\centering
\setlength{\tabcolsep}{1.0pt}
\begin{tabular}{lcccccccccccc}
\toprule 
    \textbf{\makecell[bl]{Dataset}}
    & \multicolumn{3}{c}{\textbf{\makecell[br]{Movies}}} 
    & \multicolumn{3}{c}{\textbf{\makecell[br]{Fashion}}} 
    & \multicolumn{3}{c}{\textbf{\makecell[br]{Games}}} 
    & \multicolumn{3}{c}{\textbf{\makecell[br]{Music}}} \\
    \cmidrule(lr){2-4} \cmidrule(lr){5-7} \cmidrule(lr){8-10} \cmidrule(lr){11-13}

    \quad \textbf{\makecell[bl]{Metrics}} & \textbf{\makecell[br]{\rot{Recall}}} & \textbf{\rot{NDCG}} & \textbf{\rot{MRR}} & \textbf{\rot{Recall}} & \textbf{\rot{NDCG}} & \textbf{\rot{MRR}} & \textbf{\rot{Recall}} & \textbf{\rot{NDCG}} & \textbf{\rot{MRR}} & \textbf{\rot{Recall}} & \textbf{\rot{NDCG}} & \textbf{\rot{MRR}}\\
    \midrule
        \textit{General} &&&&&&&&&& \\
            \quad PopRec              & .0387 & .0186 & .0126 & .0246 & .0104 & .0064 & .0000 & .0000 & .0000 & .0005 & .0002 & .0001 \\

            \quad BPR                 & .0704 & .0365 & .0263 & .1359 & .1303 & .1286 & .0065 & .0032 & .0022 & .0007 & .0003 & .0002 \\
            \quad NCF                & .0310 & .0147 & .0099 & .1130 & .0693 & .0557 & .0303 & .0157 & .0113 & .0007 & .0006 & .0005 \\
            \quad DiffRec & \textbf{.2459} & \textbf{.1368} & \textbf{.1037} & .1748 & .1487 & \underline{.1398} & \textbf{.1003} & \textbf{.0558} & \textbf{.0423} & \underline{.0797} & \underline{.0479} & \underline{.0382} \\
    \textit{Context-aware} &&&&&&&&&& \\
        \quad De-biased MF        & .0566 & .0271 & .0183 & .1303 & .1236 & .1216 & .0075 & .0032 & .0019 & .0027 & .0011 & .0006 \\
        \quad FM                  & .0467 & .0221 & .0149 & .1088 & .1087 & .1001 & .0089 & .0041 & .0011 & .0042 & .0020 & .0013 \\
        \quad DICE  & .1755 & .0941 & .0695 & \textbf{.1767} & \underline{.1482} & .1385 & .0677 & .0355 & .0257 & .0200 & .0113 & .0087 \\
    \textit{Sequential} &&&&&&&&&& \\
        \quad TransRec            & .0211 & .0098 & .0065 & .1464 & .1127 & .1013 & .0101 & .0038 & .0020 & .0005 & .0001 & .0000 \\
        \quad Caser               & .1601 & .0810 & .0577 & .0429 & .0172 & .0096 & .0000 & .0000 & .0000 & .0000 & .0000 & .0000 \\
        \quad GRU4Rec             & .1599 & .0794 & .0531 & .1188 & .0826 & .0713 & .0705 & .0366 & .0263 & .0184 & .0083 & .0052 \\
        \quad BERT4Rec            & .0303 & .0132 & .0082 & .1187 & .0485 & .0283 & .0006 & .0003 & .0002 & .0000 & .0000 & .0000 \\
        \quad SASRec              & .1597 & .0786 & .0542 & .1363 & .0801 & .0620 & .0523 & .0263 & .0186 & .0426 & .0251 & .0199 \\
        \quad FEARec & \underline{.2402} & \underline{.1211} & \underline{.0851} & \underline{.1762} & \textbf{.1572} & \textbf{.1508} & .0435 & .0205 & .0136 & \textbf{.1402} & \textbf{.0724} & \textbf{.0515} \\
    \textit{Sequential+} &&&&&&&&&& \\
    \textit{Context-aware}&&&&&&&&&& \\
        \quad \textbf{Ours}       & .1677 & .0853 & .0604 & .1459 & .1200 & .1112 & \underline{.0724} & \underline{.0387} & \underline{.0284} & .0682 & .0352 & .0253 \\

    \midrule
    Improvement & 4.74\% & 5.30\% & 4.68\% & -0.34\% & -7.90\% & -13.53\% & 2.70\% & 5.74\% & 7.98\% & 60.09\% & 40.23\% & 27.15\% \\
\bottomrule
\end{tabular}
\end{table}

\begin{table}[htb!]
  \centering
  \caption{Accuracy metrics over the users who prefer popular and niche items, respectively. The best-performing models are in bold, and the second-best-performing models are underlined.}
    \setlength{\tabcolsep}{1.4pt}
    \small
    \begin{tabular}{lccccccccccccccc}
    \makecell[lb]{
    \textbf{Dataset}\\
    \quad \textbf{Metrics}} &
    \makecell[rb]{\rot{RandomRec}} & 
    \makecell[rb]{\rot{PopRec}} & 
    \makecell[rb]{\rot{$\epsilon$-greedy}} &
    \makecell[rb]{\rot{BPR}} & 
    \makecell[rb]{\rot{NCF}} & 
    \makecell[rb]{\rot{DiffRec}} & 
    \makecell[rb]{\rot{De-biased MF}} &
    \makecell[rb]{\rot{FM}} &
    \makecell[rb]{\rot{DICE}} & 
    \makecell[rb]{\rot{TransRec}} &
    \makecell[rb]{\rot{GRU4Rec}} &
    \makecell[rb]{\rot{SASRec}} &
    \makecell[rb]{\rot{FEARec}} & 
    \textbf{\makecell[rb]{\rot{BiCoRec}}} &
    \makecell[rb]{\rot{Improvement}} \\
    \midrule
        \textbf{Movies} &&&&&&&&&& \\
            \quad \textbf{\textit{Popular}}  &&&&&&&&&& \\
                \quad \quad Recall@10      & .0050 & .0135 & .0081 & .0122 & .0086 & .0124 & .0008 & .0013 & .0133 & .0008 & \underline{.0106} & .0078 & \textbf{.0134} & .0073 & -45.52 \% \\
                \quad \quad NDCG@10        & .0023 & \textbf{.0075} & .0052 & .0059 & .0037 & .0055 & .0004 & .0006 & .0065 & .0004 & .0050 & .0036 & .0057 & \underline{.0067} & -10.67 \% \\
            \quad \textbf{\textit{Niche}}  &&&&&&&&&& \\
                \quad \quad Recall@10      & .0020 & .0028 & .0028 & .0052 & .0035 & .0041 & .0065 & \underline{.0088} & .0058 & .0005 & .0044 & .0017 & .0020 & \textbf{.0098} & 11.36 \% \\
                \quad \quad NDCG@10        & .0007 & .0011 & .0010 & .0029 & .0013 & .0019 & .0077 & \underline{.0094} & .0027 & .0002 & .0075 & .0053 & .0011 & \textbf{.0102} & 8.51 \% \\
     \midrule
        \textbf{Fashion}  &&&&&&&&&& \\
            \quad \textbf{\textit{Popular}}  &&&&&&&&&& \\
                \quad \quad Recall@10      & .0006 & .0394 & .0376 & .0480 & .0417 & .0273 & .0434 & .0122 & .0372 & .0457 & .0406 & \underline{.0485} & \textbf{.0502} & .0412 & -17.93 \% \\
                \quad \quad NDCG@10        & .0003 & .0129 & .0123 & \underline{.0439} & .0123 & .0247 & .0353 & .0062 & .0243 & .0281 & .0231 & .0402 & \textbf{.0445} & .0270 & -39.33 \% \\
            \quad \textbf{\textit{Niche}}  &&&&&&&&&& \\
                \quad \quad Recall@10      & .0010 & .0000 & .0005 & .0005 & .0043 & .0000 & \underline{.0075} & .0046 & .0167 & .0000 & .0000 & .0010 & .0016 & \textbf{.0098} & 30.67 \% \\
                \quad \quad NDCG@10        & .0007 & .0000 & .0002 & .0002 & .0010 & .0000 & \underline{.0039} & .0019 & .0010 & .0000 & .0000 & .0003 & .0010 & \textbf{.0047} & 20.51 \% \\
     \midrule
        \textbf{Games}  &&&&&&&&&& \\
            \quad \textbf{\textit{Popular}}  &&&&&&&&&& \\
                \quad \quad Recall@10      & .0002 & .0000 & .0000 & .0004 & \underline{.0085} & .0046 & .0008 & .0015 & .0054 & .0043 & .0052 & .0056 & .0047 & \textbf{.0106} & 11.47 \% \\
                \quad \quad NDCG@10        & .0000 & .0000 & .0000 & .0003 & \underline{.0034} & .0020 & .0004 & .0007 & .0030 & .0021 & .0030 & .0024 & .0021 & \textbf{.0057} & 67.65 \% \\
            \quad \textbf{\textit{Niche}}  &&&&&&&&&& \\
                \quad \quad Recall@10      & .0000 & .0000 & .0000 & .0000 & .0000 & .0000 & .0009 & .0012 & .0000 & \underline{.0055} & .0005 & .0000 & .0000 & \textbf{.0089} & 61.81 \% \\
                \quad \quad NDCG@10        & .0000 & .0000 & .0000 & .0000 & .0000 & .0000 & .0002 & .0009 & .0000 & \underline{.0017} & .0003 & .0000 & .0000 & \textbf{.0021} & 23.53 \% \\
     \midrule
        \textbf{Music}  &&&&&&&&&& \\
            \quad \textbf{\textit{Popular}}  &&&&&&&&&& \\
                \quad \quad Recall@10      & .0000 & .0000 & .0000 & .0000 & .0000 & .0028 & .0005 & .0011 & .0018 & .0000 & .0056 & .0083 & \textbf{.0088} & \underline{.0092} & -4.55 \% \\
                \quad \quad NDCG@10        & .0000 & .0000 & .0000 & .0000 & .0000 & .0013 & .0002 & .0005 & .0008 & .0000 & .0026 & .0039 & \textbf{.0042} & \underline{.0048} & -14.29 \% \\
            \quad \textbf{\textit{Niche}}  &&&&&&&&&& \\
                \quad \quad Recall@10      & .0000 & .0000 & .0000 & .0000 & .0000 & .0000 & .0008 & .0016 & \underline{.0017} & .0000 & .0004 & .0000 & .0017 & \textbf{.0024} & 50.00 \% \\
                \quad \quad NDCG@10        & .0000 & .0000 & .0000 & .0000 & .0000 & .0000 & .0002 & \underline{.0004} & .0003 & .0000 & .0003 & .0000 & .0002 & \textbf{.0005} & 25.00 \% \\
                
    \bottomrule
    \end{tabular}  \label{Table:PopularityMetrics}
\end{table}




\end{appendices}

\end{document}